\documentstyle[aps,prb,eqsecnum,multicol,psfig]{revtex}
\begin{document}
\draft
\title{\hfill  \\ \vspace{10pt}
Transport in a one-dimensional Luttinger liquid}
\author{Matthew P. A. Fisher$^{1}$ and Leonid I. Glazman$^{2}$  }
\address{$^{1}$Institute for Theoretical Physics, University of California,
Santa
Barbara, CA 93106--4030\\$^{2}$Theoretical Physics Institute, University of
Minnesota, Minneapolis, MN 55455}

\date{\today}
\maketitle
\psfigurepath{}

\begin{abstract}
In this paper we review recent theoretical results for transport in a
one-dimensional (1d) Luttinger liquid. For simplicity, we ignore electron
spin, and focus exclusively on the case of a single-mode. Moreover, we
consider only the effects of a single (or perhaps several) spatially
localized impurities. Even with these restrictions, the predicted behavior
is very rich, and strikingly different than for a 1d non-interacting
electron gas. The method of bosonization is reviewed, with an emphasis on
physical motivation, rather than mathematical rigor. Transport through a
single impurity is reviewed from several different perspectives, as a pinned
strongly interacting ``Wigner" crystal and in the limit of weak interactions.
The existence of fractionally charged quasiparticles is also revealed.
Inter-edge tunnelling in the quantum Hall effect, and charge fluctuations in
a quantum dot under the conditions of Coulomb blockade are considered as
examples of the developed techniques.
\end{abstract}

\begin{multicols}{2}
\section{Introduction}
\label{Introduction}
Landau's Fermi liquid theory is a beautiful and successful
theory which accounts for the behavior of the conduction electrons
in conventional metallic systems\cite{AGD}.
The central assumption of this theory, is that the
low energy excited states of the
interacting electron gas can be classified in the same way as a
reference
non-interacting electron gas.
Upon adiabatically ``switching on" the electron interactions,
the excited states are still
composed of free quasiparticles,
labelled by momentum,
which obey Fermi statistics.
The low energy excitations above the ground state, consist of
creating quasiparticles (and quasiholes) with momenta just above (below)
the Fermi surface.
The
single-particle spectral function for an electron is
presumed to have a $\delta$-function peak at the Fermi surface,
${{\cal A}}(k_F,\epsilon)=2\pi Z_F \delta (\epsilon)$,
with a non-zero coefficient, $Z_F$.
For the non-interacting electron gas $Z_F=1$,
but with interactions $Z_F<1$, reflecting a smaller overlap
between the electron and the quasiparticle.
However, provided $Z_F$ is {\it non-zero},
there exist well defined low energy quasiparticle excitations,
which are in a one-to-one
correspondence with the bare electron excitations
of the reference non-interacting electron gas.
It is then possible to build a transport theory
of Fermi liquids in direct analogy with Drude theory
of the free electron gas, by focussing on the scattering
(by impurities, say) of the low energy quasiparticle
excitations.
For example, a quasiparticle will scatter
off an impurity with a finite cross-section,
leading to a metallic resisitivity
for low impurity
concentration.

For a weakly-interacting electron gas,
deviations in the spectral weight $Z_F$ from one can be computed
perturbatively in the interaction strength,
provided the spatial dimensionality, $d$, is greater than one.
However, for the
one-dimensional interacting system, low order terms in the
perturbation theory are
divergent.  This signals the breakdown of Landau's Fermi liquid
theory in the 1d electron gas, for {\it arbitrarily} weak
interactions\cite{Shankar}.
This divergence can be traced to
the large phase space in 1d, for
an electron to relax by creating
an electron-hole excitation.
For $d>1$,
both energy and momentum conservation highly constrain the available
phase space, but in 1d
for a single branch with linear dispersion, momentum
conservation automatically implies energy
conservation, so the phase space is less constrained.
This leads to a
divergent rate for such scattering processes,
and the electron becomes ``dressed" by a
large number of electron-hole pairs.
This in turn drives the electron spectral weight to zero,
$Z_F=0$, even for weak electron interactions.
In the limit of weak interaction, it is still
possible to extract the resulting properties of the 1d
electron gas by a clever resummation of the
most divergent terms in each order of perturbation theory\cite{Solyom}.
However, the physics is more readily revealed
by the method of bosonization, as discussed below.

Tomonaga\cite{Tomonaga} (and more recently Luttinger\cite{Luttinger})
considered
a special class of interacting 1d electron models, which were linearized around
the Fermi points, and could be diagonalized in terms of boson variables.  These
boson variables described collective plasmon excitations,  or density waves, in
the electron gas.   Some years later, Haldane\cite{Haldane} argued that this
bosonized decription was valid more generally, giving an appropriate description
for the low energy excitations of a generic 1d interacting electron gas. Haldane
coined the term ``Luttinger liquid" to describe this generic behavior, although
sometimes it is referred to as a Tomonaga-Luttinger liquid. In such a 1d
Luttinger liquid, creation of a real electron is achieved by exciting an
infinite number of plasmons.  Because of this, the space and time dependence of
the electron correlation function is dramatically different than in a
non-interacting electron gas, or in a Fermi liquid\cite{Mahan}. This manifests
itself in various kinetic quantities, in particular the Drude conductivity,
which is predicted to vary (as a power law) with
temperature\cite{Mattis,luther1,luther2}. The Luttinger liquid approach to
transport  in 1d systems was popular in the 1970`s, being employed in attempts
to understand the behavior of quasi-1d organic conductors.

Recent advances in semiconductor technology have renewed interest in
transport of 1d electron systems.
By cleverly ``gating" a
high mobility two-dimensional (2d) electron gas,
it is possible to further confine the motion of the electrons,
so that they can only move freely along a 1d channel.
Moreover, it is possible to make a very narrow channel, with width
comparable to the electron`s Fermi wavelength.
In this case, the
transverse degrees of freedom are quantized,
and only one, or perhaps several, 1d channels
are occupied at the Fermi level.
This ``quantum wire" provides an experimental realization of a 1d electron gas,
which should reveal the signatures of Luttinger liquid behavior without
the complicated crossovers to three-dimensional behavior
inherent in the quasi-1d organic conductors.

Edge states formed at the boundaries of the 2d electron gas
when placed in strong magnetic field in the quantum Hall regime,
provide another important example of a one-dimensional electron
system\cite{Halperin}.  These edge excitations
are chiral, moving only in one direction along the edge.
For a quantum Hall bar geometry, the two edge modes confined
to the edges move in opposite directions.
Together they comprise a non-chiral system, which
resembles a 1d electron gas.
For the integer quantum Hall effect, these edge excitations are
equivalent to a 1d non-interacting electron gas,
(despite the presence of electron interactions).
However, in the {\it fractional} quantum Hall effect,
the edge excitations are believed to be isomorphic
to a 1d interacting Luttinger liquid\cite{Wen}.  Moreover, due
to the spatial separation between the modes on opposite edges,
impurities cannot backscatter, so that localization effects
important in a quantum wire
are absent here.

A 1d quantum wire is appropriately characterized by a conductance,
which can be measured in a transport experiment.
In the absence of interactions, the conductance of an
ideal single-mode channel wire,
adiabatically connected\cite{adiabatic} to leads, is quantized, $G= 2 e^2/h$
-- where the factor of 2 accounts for spin.
If a scatterer
is introduced into the channel, the conductance drops,
with $G =2{\cal T}e^2/h$-- where ${{\cal
T}}<1$ is
the transmission coefficient for electrons at the Fermi level.
For edge states in the integer quantum Hall effect,
a quantized conductance occurs
when there is no backscattering between
modes on opposite edges, and corresponds to
the quantized Hall conductance.
Inter-edge scattering is equivalent to a reduced transmission
coefficient, and reduces the conductance.

As discussed above,
electron interactions should dramatically
modify the low energy excitations in the quantum wire.
As we shall see, this leads to striking predictions
for the transport
in a quantum wire with one or several impurities.
Likewise, inter-edge
backscattering in the fractional quantum Hall effect
is predicted to be very different than in the integer Hall effect.

In this paper we review recent theoretical results for
transport in a 1d Luttinger liquid.
For simplicity, we ignore electron spin, and
focus exclusively on the case of a single-mode.
Moreover, we consider only the effects of a single
(or perhaps several) spatially localized impurities.
As we shall see, even with these restrictions, the
predicted behavior is very rich, and strikingly different
than for a 1d non-interacting electron gas.

The paper is organized as follows.
In Section II the method of bosonization is reviewed, with an emphasis
on physical motivation, rather than mathematical rigor.
After discussing briefly the conductance of an ideal
channel, we reveal the existence of fractionally charged
excitations in
the Luttinger liquid.  These correspond to Laughlin quasiparticles
for fractional quantum Hall edge states, but should also
be present in quantum wires.
In Section
\ref{tunneling}, we consider the tunnelling density of states
for adding an electron into a 1d Luttinger liquid, which
is suppressed, vanishing as a power law of energy.
The exponent is extracted for
both tunnelling into the middle of an infinite
1d wire, and into the ``end" of a semi-infinite wire.
Section \ref{transport} is devoted to
a detailed analysis of transport in a Luttinger liquid with a single
barrier or impurity.
We consider two limiting cases, a very large barrier in
IVA and a very small barrier in IVB.  In Section IVC we show how the
crossover between these two limits can be understood by considering
general barrier strengths, but weak electron interactions.  A general picture
of this crossover is described in IVD, and the special case
of resonant tunnelling is discussed in IVE.

In Section V, we consider briefly two particular applications.
In VA we consider tunnelling between edge states in the fractional
quantum Hall effect,
which is a rather straightforward application of
the general theory.  In Section VB we employ Luttinger liquid theory
to describe Coulomb blockade in a quantum dot.
Section VI is devoted to a very brief summary, and list
of other related topics.

\section{Bosonization}
\label{bosonization}

Consider the Hamiltonian for spinless electrons hopping on a 1d lattice,
\begin{equation}
{\cal H} = -t \sum_j c_j^\dagger c_{j+1} + h.c. + {\frac{V}{2}} \sum_j
c_j^\dagger c_j
c_{j+1}^\dagger c_{j+1} ,
\label{Hamiltonian}
\end{equation}
with hopping strength $t$ and near-neighbor interaction strength $V$. When
$V=0$ one can diagonalize the problem in terms of plane waves with energy
$E_k = -t \cos (k)$ for
momentum $k$ satisfying $|k| < \pi$. The low energy excitations consist of
particle/hole excitations
across the two Fermi points, at $\pm k_F$. Consider a single particle
hole/excitation about the right
Fermi point, where a single electron is removed from a state with $k<k_F$
and placed into an unoccupied
state with $k+q>k_F$. For small momentum change $q$, the energy of this
excitation is
$\omega_k = v_F q$, with $v_F$ the Fermi velocity. Together with the
negative momentum excitations
about the left Fermi point, this linear dispersion relation is identical to
that for phonons in
one-dimension. The method of Bosonization exploits this similarity by
introducing a phonon
displacement field, $\theta$ to decribe this linearly dispersing density
wave. When interactions are
turned on, the single particle nature of the excitations is not retained,
but this linearly dispersing mode
survives, with a renormalized velocity.

The method of Bosonization focusses on the low energy excitations, and provides
an effective theory. We follow the heuristic development of
Haldane\cite{Haldane}, which reveals the important physics, dispensing with
mathematical rigor. To this end, consider a Jordan-Wigner transformation which
replaces the electron operator, $c_j$, by a (hard-core) boson operator,
$c_j = \exp (i\pi \sum_{j < i} n_i) b_j$, where $n_j = c^\dagger_j c_j$ is the
number operator. One can easily verify that the Bose operators commute at
different sites. Moreover, the above Hamiltonian when re-written is of the
identical form with $c's$ replaced by $b's$. This transformation, exchanging
Fermions for Bosons, is a special feature of one-dimension. The Boson operators
can be (approximately) decomposed in terms of an amplitude and a phase, $b
\rightarrow \sqrt{n_j}\exp (i\phi_j)$. We now imagine passing to the continuum
limit, focussing on scales long compared to the lattice spacing. In this limit
we replace, $\phi_j\rightarrow\phi(x)$ and $n_j \rightarrow\rho(x)$, where
$\rho(x)$ is the 1d electron density. This can be decomposed as
$\rho = \rho_0 + \tilde{\rho}$, where the mean density, $\rho_0 = k_F/\pi$, and
$\tilde{\rho}$ is an operator measuring fluctuations in the density. As usual,
the density and phase are canonically conjugate quantum variables, taken to
satisfy
\begin{equation}
[\phi(x), \tilde{\rho}(x')] = i \delta(x-x') .
\label{commutation}
\end{equation}

Now we introduce a phonon-like displacement field, $\theta(x)$, via
$\tilde{\rho}(x) =\partial_x \theta(x) /\pi$. The factor of $\pi$ has been
chosen so that the full density takes the simple form: $\rho(x) = (k_F +
\partial_x \theta)/\pi$. The above commutation relations are satisfied if one
takes,
\begin{equation}
[\phi(x), \theta(x')] = {{i\pi} \over {2}} {\rm sgn}(x-x') .
\label{commutation1}
\end{equation}
Notice that $\partial_x \phi$ is the momentum conjugate to $\theta$. The
effective (Bosonized) Hamiltonian density which describes the 1d density wave
takes the form:
\begin{equation}
H = {v \over {2\pi}} [ g (\partial_x \phi)^2 + g^{-1} (\partial_x \theta)^2] .
\label{Hamiltonian1}
\end{equation}
This Hamiltonian describes a wave propagating at velocity $v$, as can be readily
verified upon using the commutation relations to obtain the equations of motion,
$\partial_t^2\theta = v^2
\partial_x^2 \theta$, and similarly for $\phi$. For non-interacting electrons
one should equate
$v$ with the (bare) Fermi velocity, $v_F$, but with interactions $v$ will be
modified (see below). The additional dimensionless parameter, $g$, in the above
Hamiltonian also depends on the interaction strength. For non-interacting
electrons one can argue that $g=1$, as follows. A small variation in density,
$\tilde{\rho}$ will lead to a change in energy, $E =\tilde{\rho}^2/2\kappa$,
where $\kappa = \partial\rho/\partial \mu$ is the compressibility. Since
$\partial_x\theta = \pi\tilde{\rho}$, one deduces from $H$ that
$\kappa = g/\pi v$. But for a non-interacting electron gas, $\pi v \kappa = 1$,
so that $g=1$.

To see why $g$ shifts with interactions, it is convenient first to identify the
electron creation operator. On physical grounds, the appropriate operator,
denoted $\psi^\dagger (x)$, should create a unit charge ($e$) excitation at
point $x$, which anti-commutes with the same operator at a different point $x'$.
Consider the Bose creation operator, $b^\dagger \sim \exp (i\phi)$. To see that
this indeed creates a unit charge ($e$) excitation, one can write,
\begin{equation}
e^{i\phi(x)} = e^{i \pi \int_{- \infty}^x dx' P(x')} ,
\label{shiftoperator}
\end{equation}
where $P =\partial_x \phi/\pi$ is the momentum conjugate to $\theta$. Since
the momentum operator is
the generator of translations (in $\theta$), this creates a kink in
$\theta$ of height $\pi$ centered at
position $x$. This corresponds to a localized unit of charge, since the
density $\tilde{\rho} =\partial_x
\theta /\pi$.

To construct an electron operator requires multiplying the Bose operator
$e^{i\phi}$ by a Jordan-Wigner ``string": $e^{i\pi \sum_{j<i} n_i}
\rightarrow e^{i \pi
\int^x \rho}$. Rewriting in terms of $\theta$, this corresponds to
$e^{i(k_F x + \theta)}$. The most
general Fermionic charge $e$ operator can thus be written,
\begin{equation}
\psi(x) = \sum_{m_{\rm odd}} \psi_m \approx \sum_{m_{\rm odd}} e^{im(k_F x
+ \theta(x))} e^{i\phi(x)} .
\label{bosons}
\end{equation}
The requirement that the integer $m$ is odd is dictated by Fermi
statistics, since one can show using the
commutation relations, that
\begin{equation}
\psi_m(x) \psi_{m'}(x') = e^{i\pi (m - m') {\rm sgn}(x-x')} \psi_{m'}(x')
\psi_m(x) .
\label{comm}
\end{equation}
The terms with $m = \pm 1$ have a very simple interpretation. The electron
field operator has been expanded into a right and left moving piece,
corresponding to the right and
left Fermi points at $\pm k_F$:
\begin{equation}
\psi(x) \approx \psi_R + \psi_L \approx e^{ik_Fx} e^{i\Phi_R} + e^{-ik_Fx}
e^{i\Phi_L} ,
\label{smoothpart}
\end{equation}
with the definition, $\Phi_{R/L} = \phi \pm \theta$. Here $\Phi_{R/L}$
describe the slowly varing piece
of the electron field. These two fields commute with one another, and satisfy:
\begin{equation}
[\Phi_R(x),\Phi_R(x')]=-[\Phi_L(x),\Phi_L(x')]= i\pi {\rm sgn}(x-x') .
\label{commutation2}
\end{equation}
Notice that these unusual commutation relations - referred to as Kac-Moody
- imply that the field
conjugate to $\Phi$ is the derivative of $\Phi$ itself. These fields are
simply related to the right and left
moving electron densities, denoted $N_{R/L}$, via
\begin{equation}
N_{R/L} = \pm {1 \over {2\pi}} \partial_x \Phi_{R/L} .
\label{leftrightmovers}
\end{equation} Note that
these densities sum to give the total density, $N_R + N_L = \tilde{\rho}$.

It is instructive to re-write the Hamiltonian (\ref{Hamiltonian1})  in
terms of the right and left electron
densities,
\begin{equation} H = \pi v_0 [N_R^2 + N_L^2 + 2\lambda N_R N_L ] ,
\label{Hamiltonian2}
\end{equation}
with
\begin{eqnarray}
v_0 =v(g+g^{-1})/2
\nonumber\\
\lambda = (1-g^2)/(1+g^2).
\label{parameters}
\end{eqnarray}
This Hamiltonian descibes the system of
right- and left-moving electrons with the interaction strength
$\lambda$ between the two species. Notice that for $g=1$ the interaction
vanishes, and $v_0=v$;
Hamiltonian (\ref{Hamiltonian2}) then gives the Bosonized decsription of
the non-interacting electron
gas, and $v_0\equiv v_F$, the bare Fermi velocity. Repulsive interactions,
($\lambda > 0$) correspond to
$g<1$, whereas attractive interactions give $g>1$.

Since Hamiltonians (\ref{Hamiltonian1}) and (\ref{Hamiltonian2}) give only
an effective low energy
description, it is generally difficult to relate the velocity $v$,
and the interaction parameter
$g$, to the bare interaction strength in the
original lattice model, say
Eqn.~(\ref{Hamiltonian1}).  However, for a model with
long-range electron-electron
interactions (Tomonaga model, see {\sl e.g.}, Ref.~[\onlinecite{Mahan}])
these relations can be obtained
analytically. In the Tomonaga model, only Fourier components with $k\ll
k_F$ of the interaction
potential $V(k)$ are taken into account, and the parameters in
Eqs.~(\ref{Hamiltonian1}) and
(\ref{Hamiltonian2}) can be expressed\cite{Mahan} in terms of $V(0)$. We
consider here a generalized Tomonaga model, which allows for different values of
interaction between the electrons moving in the same direction
($V_{ll}(0)=V_{rr}(0)\equiv V_1$) and in
opposite directions ($V_{lr}(0)=V_{rl}(0)\equiv V_2$).
The velocity $v$ and interaction parameter $g$ can be related
to $V_1$ and $V_2$ as follows.
In the generalized Tomonaga model,
the
energy associated with a
density fluctuation in the right- or left-moving mode ($N_R$ or $N_L$) is a sum
of two terms, representing
compressibility of the non-interacting Fermi gas, and the interaction
potential $V_1$, respectively.
Equating this with the coefficient of $N_R^2$ in \ref{Hamiltonian2},
gives: $\pi
v_0=\pi v_F + V_1$.
The interaction between the modes $V_2$ is responsible
for the last term in
Eq.~(\ref{Hamiltonian2}), so that $\pi v_0\lambda=V_2$.
These two relations can be used to solve for the
velocity $v$ and parameter $g$, giving\cite{Dzyaloshinskii}
\begin{equation}
v=v_F\left(1+\frac{V_1}{\pi v_F}+\frac{V_1^2-V_2^2}{4\pi^2
v_F^2}\right)^{1/2} ,
\end{equation}
\begin{equation}
g=\left(\frac{1+V_1/2\pi v_F-V_2/2\pi v_F}{1+V_1/2\pi v_F+V_2/2\pi
v_F}\right)^{1/2}.
\label{Tomonaga}
\end{equation}
In the conventional case $V_1=V_2>0$, and $v$ is simply the
plasmon velocity, increasing above $v_F$ with repulsive interactions which
reduce the compressibility of the electron gas.
Notice, moreover, that the sign of $1-g$
is determined by the sign of
the interaction, with $g<1$ for repulsive interactions.
Edge states propogating on opposite
edges of an integer quantum Hall bar, constitute
a 1d system with $V_1>V_2$.
In particular, for a wide Hall bar, the inter-edge interaction
$V_2$ vanishes,
so that $g=1$ .

\subsection{Meaning of g}

No matter what the underlying microscopic model is, the parameter $g$ of
the Luttinger liquid
(\ref{Hamiltonian1}) can be given a simple physical meaning as a
dimensionless conductance. To this end
we define new variables $\phi_{R/L} = g\phi \pm \theta$ which decouple the
right and left moving
sectors, and at the same time diagonalize the Hamiltonian
(\ref{Hamiltonian1}). The corresponding right
and left moving densities are
\begin{equation}
n_{R/L} =\pm {1\over {2\pi}} \partial_x\phi_{R/L} .
\label{leftright}
\end{equation}
Although $n_{R/L}$ again sum to give the total density, $n_R + n_L =
\tilde{\rho}$, for $g\ne 1$ they do not correspond to the right and left
pieces of an {\it electron}, but
rather involve contributions from electrons propagating in both directions.
The right and left fields
commute with one another and satisfy,
\begin{equation}
 [\phi_R(x),\phi_R(x')]=-[\phi_L(x),\phi_L(x')]= i\pi g {\rm sgn}(x-x') ,
\label{commutation3}
\end{equation}
the same as for $N_{R/L}$ except with an important factor of $g$ on the
right side. In terms of these
fields, the Bosonized Hamiltonian decouples into right and left moving sectors:
\begin{equation}
H = {{\pi v} \over g} [n_R^2 + n_L^2 ] .
\label{lrhamiltonian}
\end{equation}
This innocent looking decoupling, leads to a number of remarkable
predictions for the 1d electron gas, as
discussed below. Firstly, from the Hamiltonian one can generate the
equations of motion, which are
\begin{equation}
(\partial_t \pm v \partial_x )n_{R/L} = 0 .
\label{lrequation}
\end{equation}
These chiral wave equations have the general solutions, $n_{R/L} = f(x \mp
vt)$, for arbitrary
$f$, and describe a density disturbance which propagates, without
scattering (or dispersion), to the right
or left. This, despite the fact that the electrons, which are interacting,
do scatter off one another.

The physical meaning of $g$ as a conductance can now be revealed. Imagine
an external probe (or
contact) which raises the chemical potential of the chiral mode $n_R$, by
an amount $\mu_R$. This can
be incorporated by adding to the Hamiltonian density a term of the form,
$\delta H = - e \mu_R n_R$. This leads to a shift in the right moving
density, since the total Hamiltonian
is no longer minimized by $n_R=0$. Rather, minimization with respect to
$n_R$, gives
$n_R = (ge/2\pi v) \mu_R$. This extra density carries an additional
(transport) current to the right, of
magnitude, $I_R = e n_R v$. One thus has, $I_R = G \mu$ with a conductance,
\begin{equation}
G = g {{e^2} \over h} .
\label{Landauer}
\end{equation}
For non-interacting electrons ($g=1$), we recover the result of Landauer
transport theory -- $e^2/h$
conductance per channel. But with interactions $G$ is modified.

Unfortunately, in a quantum wire, it is essentially impossible to
selectively couple to the right moving
mode $n_R$, and not the left. Rather, in a typical transport experiment,
the quantum wire is connected
at its ends to bulk metallic contacts, which can be modelled as free
electron gases (Fermi-liquids). In this
case, the dc transport will be sensitive to the contacts.
This can be seen by considering an alternative derivation
of
Eq. (\ref{Landauer}), based on linear response theory. Consider an
ac electric field
$E(x,t)=-U'(x)\sin\omega t$, applied along an infinite wire,
which is non-zero for $x$ near the origin, say $|x|<L/2$.
This leads to a term of
the form $geE(x,t)$ on the right hand side of
(\ref{lrequation}).
This field will cause an excitation of density
waves $n_{R/L}(x,t)$, which propogate along the wire. For
$\omega < v/L$,
the wavelength of the excitation becomes much larger than
$L$, so that one may then substitute
$U'(x)=-U\delta (x)$, with $U$ the
amplitude of the applied ac voltage across
the length $L$ segment. The equations of motion can now be
readily solved, giving for the
current
response, $I(x,t)=ge^2U\cos\omega [t-x sgn(x)/v]$.
Although the current depends on $x$, this dependence vanishes
in the $\omega\to 0$ limit,
and gives a dc conductance $G=I/U = ge^2/h$,
which reproduces Eq. (\ref{Landauer}).
The
conductance is clearly related to the
energy radiated away from the length $L$ segment.
Indeed, the radiated energy per unit time is exactly equal to
the power $U^2G/2$ absorbed from the field.
The resulting conductance is finite, even in the
absence of any disorder.
It is clear from this arguemnt that the dc conductance,
proportional to the power radiated away, depends on
the value of $g$ in the segment of wire {\it away} from the length
$L$ region.  For a wire of length $L$ attached
(adiabatically) to Fermi liquid leads, one can take
$g=1$ in the leads\cite{Maslov}, and one then expects a dc conductance
given by $G=e^2/h$.  The value of $g \ne 1$ in the wire,
can be revealed by an a.c. conductance measurement
for frequencies $\omega >v/L$, which has magnitude\cite{Kane}\cite{Matveev}
$G(\omega)=ge^2/h$.

There is a physical system however,  in which the right and left modes,
$n_{R/L}$ can be coupled to selectively. As discussed in more detail in
Section~\ref{edgestates}
below, in the fractional quantum Hall effect (FQHE) at such filling factors
$\nu<1$, that
$\nu^{-1}$ is an odd integer, a single current carrying mode is predicted
along the sample edges.
In a Hall bar geometry, the right and left moving modes are localized on
the top and bottom
edges of the sample, respectively. These modes are mathematically
isormorphic to the right and
left moving modes of the 1d interacting electron gas,
$n_{R/L}$, provided one makes the identification $g = \nu$. In the FQHE
geometry, a chemical potential
difference between the right and left moving modes, corresponds to a Hall
potential drop, transverse to
the transport current. The above conductance, corresponds directly to the
quantized Hall conductance, $G
=\nu e^2/h$.

\subsection{Charge fractionalization}

The above analogy with the FQHE, suggests the possibility of fractionally
charged excitations in the 1d
electron gas. To see these, it is convenient to consider adding a small
impurity potential, localized at the
origin $x=0$. This potential will cause scattering between the right and
left moving modes, $n_{R/L}$.
We shall now show that the ``particle" which is backscattered is {\it not}
a (bare) electron. Rather, the
total backscattered charge is non-integral, with magnitude $ge$. For
repulsive interactions $g <1$, this
corresponds to the backscatering of a quasiparticle with fractional
electron charge. We shall also show,
that this fractionally charged quasiparticle, also has fractional
statistics. A localized impurity potential,
$U(x) = u\delta(x)$, couples to the local electron density,
$\psi^\dagger(0) \psi(0)$. Inserting
the expression (\ref{smoothpart}), gives the dominant $2k_F$ backscattering
contribution:
\begin{equation}
H_{\rm imp} = u \delta(x) (\psi^\dagger_R \psi_L + h.c.) = u\delta(x)
(e^{i2\theta} + h.c.) .
\label{Himp}
\end{equation}
Although, this perturbation backscatters a (bare) electron, there is a
``backflow" term due to the
electron-interactions. To reveal this, one should re-express the above
perturbation in terms of the fields,
$\phi_{R/L}$, which propogate freely (to the right or left) even in the
presence of electron interactions.
One gets simply,
\begin{equation} H_{\rm imp} =u \delta(x) (e^{i\phi_R}e^{-i\phi_L} + h.c.) ,
\label{Himpurity}
\end{equation}
so that the impurity hops an $e^{i\phi_{R/L}}$ quasiparticle between the
two modes. The charge created
by the operator $e^{i\phi_R}$ can be deduced from the conjugate momentum,
$P_R = \partial_x
\phi_R/2\pi g$ which follows from the commutation relations
(\ref{commutation3}). One has,
\begin{equation}
e^{i\phi_R(x)} = e^{i2\pi g \int^x P_R(x')} ,
\end{equation}
which creates a kink in $\phi_R$ of magnitude $2\pi g$ centered at $x$.
Since $n_R=\partial_x
\phi_R/2\pi$ the charge associated with this (kink) excitation is
fractional, $Q= ge$.

Thus, a localized impurity potential in an interacting 1d electron gas,
causes backscattering of
fractionally charged quasiparticles ($Q=ge$) between the free-streaming
right and left moving modes,
$n_{R/L}$. In the FQHE, a localized impurity is equivalent to a ``point
contact" in which the right and left
moving edge modes on opposite sides of the Hall bar are ``pinched"
together, to allow for inter-mode
backscattering. In this case, these backscattered quasiparticles have a
natural physical interpretation as
Laughlin quasiparticles with charge $Q =\nu e$. The fractional charge can
perhaps be revealed
in a shot noise type experiment, as discussed in Ref.~\onlinecite{noise}.

Not surprisingly, these quasiparticle excitations also have fractional
statistics. To see this we can use the
commutation relations (\ref{commutation3}) to show that,
\begin{equation}
e^{i\phi_R(x)} e^{i\phi_R(x')} = e^{i\pi g {\rm sgn}(x-x')}
e^{i\phi_R(x')}e^{i\phi_R(x)} ,
\end{equation}
so that under exchange they pick up a phase factor $\exp(i\pi g)$. For the
non-interacting electron gas,
$g=1$, the quasiparticles have charge $e$ and Fermi statistics (the
electron!), but with interactions will
have fractional statistics.

\section{Tunnelling into a Luttinger liquid}
\label{tunneling}
As discussed in the previous section, elementary excitations in the
Luttinger liquid are significantly
different from bare electrons. This difference reveals itself
in the behavior of the density of states
for tunnelling an electron into the Luttinger liquid.
In subsection A below we will consider
specifically the
local tunnelling density of states for an infinitely
long Luttinger liquid.
Another situation of interest, analyzed in subsection B,
consists of tunnelling an electron through a large barrier
separating two semi-infinite systems.
This process is sensitive to the density of states
at the ``end" of a semi-infinite Luttinger liquid,
which, in contrast to the non-interacting electron gas,
behaves differently than the ``bulk" density of states.

Since the effective Hamiltonian density (\ref{Hamiltonian1}) is quadratic
in the boson fields, local
density of states can be readily extracted. Perhaps the simplest way to
proceed, is to represent the
partition function,
\begin{equation}
Z={\rm Tr}\exp (-\beta \int_x H) ,
\end{equation}
as an imaginary time path integral over the boson fields:
\begin{equation}
Z = \int D\phi D\theta \exp (-S) ,
\label{action}
\end{equation}
where the integration is over classical fields $\phi(x,\tau)$, and similarly
$\theta$, with imaginary time $\tau$ running from $0$ to $\beta$. The
(Euclidian) action can be written
in terms of the Lagrangian, $S = \int dx d\tau {\cal L}_0$, with
\begin{equation}
{\cal L}_0 ={i \over \pi} \partial_\tau \phi \partial_x \theta +
H(\phi,\theta)	,
\label{Lagrangian0}
\end{equation}
with $H$ given in (\ref{Hamiltonian1}). It is instructive to perform the
(Gaussian) functional integration
over, say $\phi$, which gives:
\begin{equation}
{\cal L}_0^{\theta} = {1 \over {2\pi g}} [v(\partial_x \theta)^2 + v^{-1}
(\partial_\tau
\theta)^2 ] ,
\label{Lagrangiantheta}
\end{equation}
describing, via the ``displacement" field $\theta$, 1d phonons propagating
with velocity $v$. Likewise,
integration over $\theta$ gives:
\begin{equation}
{\cal L}_0^{\phi} = {g \over{2\pi}} [v(\partial_x \phi)^2 + v^{-1}
(\partial_\tau \phi)^2 ] .
\label{Lagrangianphi}
\end{equation}
This can be interpreted as a wave propagating in the phase-field $\phi$.
Notice that
$g \rightarrow g^{-1}$ upon transforming between the two representations,
$\theta \rightarrow\phi$. At the special non-interacting point, $g=1$,
there is a self-duality between
these two representations. Physically, for strong repulsive interactions
corresponding to $g <<1$,
(zero-point) fluctuations in the displacement field $\theta$ are greatly
suppressed. In this limit
$\theta$ becomes a ``good" classical variable, and the 1d electron system
is well described as a ``solid" -
or equvalently a Wigner crystal. In the opposite extreme of very strong
attractive interactions, $g>>1$,
fluctuations in the phase $\phi$ are strongly suppressed. The system is
well described as a (almost)
condensed superfluid of the (Jordan-Wigner) bosons,
$b$.

In the presence of a single impurity, the Hamiltonian $H(\phi,\theta)$ in
Eq.~(\ref{Lagrangian0}) should
be replaced by $H+H_{\rm imp}$. Using Eq.~(\ref{Himp}) and integrating out
the field $\phi$, we find
\begin{equation}
{\cal L}^{\theta} = {1 \over {2\pi g}} [v(\partial_x \theta)^2 + v^{-1}
(\partial_\tau
\theta)^2 ]+u\delta (x)(e^{2i\theta}+h.c.) .
\label{Limp}
\end{equation}
In the language of 1d
displacements, the impurity plays the role of a
pinning center, which favors a periodic set of values of the displacement
field at the pinning site. A
strong center (large $|u|$) effectively cuts the infinite system in two
semi-infinite pieces with
independent excitation spectra. Each piece is then described by the Lagrangian
(\ref{Lagrangiantheta}),
defined on the appropriate semi-infinite space with boundary condition
$\theta(x=0)=0$.

\subsection{Tunneling into a clean Luttinger liquid}
Consider now the local electron tunnelling density of states for adding an
electron at energy $E$,
\begin{equation}
\rho(E) = 2\pi \sum_n |<n|\psi^\dagger(x)|0>|^2 \delta(E_n - E_0 - E).
\label{tunndos}
\end{equation}
Here $|n\rangle$ are exact eigenstates of the full interacting
Hamiltonian, and $E_n$ are the
corresponding energies. The summation in Eq.~(\ref{tunndos}) is performed
over a complete
set of states, and therefore with the help of the identity
\begin{equation}
\delta(E_n - E_0 - E)=\frac{1}{\pi}{\rm Re}\int_0^{\infty}dte^{i(E + E_0 -
E_n)t},
\end{equation}
can be related to the electron Green's
function,
\begin{equation}
\rho(E) = \frac{1}{\pi}{\rm Re}\int_0^\infty dt e^{iEt} \langle\psi(x,t)
\psi^\dagger(x,0)\rangle .
\label{rho}
\end{equation}
This can be evaluated by computing the imaginary time correlator,
\begin{equation}
{\cal G}(\tau)= \langle T_\tau\psi(\tau) \psi^\dagger(0)\rangle ,
\label{imtime}
\end{equation}
and then performing an analytic continuation,
$\langle\psi(t)\psi^\dagger(0)\rangle = {\cal G}(\tau \rightarrow it)$.

Upon re-expressing the electron operator in terms of the boson fields,
using (\ref{smoothpart}), the
average in (\ref{imtime}) can be readily performed for a system with a
quadratic Lagrangian. For a
clean infinite 1d system, one can use the Lagrangian (\ref{Lagrangian0}).
With a frequency cutoff, $\tau_c^{-1}$, one finds,
\begin{equation}
{\cal G}(\tau) = \left( {\tau_c\over {|\tau| + \tau_c}} \right)^\alpha ,
\end{equation}
with exponent $\alpha = (g+g^{-1})/2$. After analytic continuation and Fourier
transformation, one thereby obtains,
\begin{equation}
\rho(E) = {{2\pi} \over {\Gamma(\alpha)}} \Theta(E) \tau_c^\alpha E^{\alpha-1} .
\label{tunndos1}
\end{equation}

Notice that for non-interacting electrons ($g=1$), one recovers the
expected constant density of states.
However, with interactions present, $g \ne 1$, the electron tunnelling
density of states vanishes as $E
\rightarrow 0$. This striking feature of the 1d electron gas is intimately
related to the fractional charge.
For $g \ne 1$ the excitations of the system are not free electron like, and
there is an orthogonality
catastrophe when one tries to add in an electron.  The orthogonality
catastrophe occurs because the accomodation of an added
electron requires
modification of the wavefunctions of all the electrons forming the liquid.

The tunneling conductance through a large barrier separating
two semi-infinite Luttinger liquids,
is proportional to
the product of the ``end" density of states for the two pieces.
This can be evaluated
with the help of the
Lagrangian (\ref{Lagrangianphi})
defined, say, for $x>0$,
together with the boundary condition $\theta(x=0)=0$. The result
is similar to
Eq.~(\ref{tunndos1}) with $\alpha=1/g$.
This result can be understood in rather simpler physical terms
in the limit of $g\ll 1$, as we now discuss.

\subsection{Tunneling through a barrier in the limit $g\ll 1$}

In the limit of strong electron-electron interactions, the fluctuations in
the ``displacement'' field are
small, and the 1d system with a single barrier can be treated as a Wigner
crystal pinned by an impurity.
At low energies $E$, the process of tunneling consists of several stages.

If the barrier created by the impurity is narrow and high enough,
tunneling of the electrons close to the
barrier occur on a fast time scale, set by the Wigner crystal
Debye frequency,
$\omega_D\equiv k_Fv$. This fast process results in a sudden
($|t|\lesssim 1/\omega_D$)
creation of an electron vacancy--interstitial pair separated by the barrier. The
energy of such a configuration is
large, $\sim\omega_D/g$, and therefore the corresponding state of the
Wigner crystal is classically forbidden.
Gradual relaxation of the electron localized
near the barrier, constitutes the slow stage of
tunneling.

Complete relaxation requires that all the electrons of the Wigner crystal
shift from their initial
positions by one crystalline period. Since the overlap
between the initial and
final wave functions for {\em each} electron is suppressed,
in the thermodynamic limit the initial and final many-body
ground states are orthogonal to
one other - the orthogonality catastrophe. This implies that
the tunneling amplitude at $E=0$ vanishes. At $E>0$, a
complete relaxation
is not required, since the system can end up in an excited state.
The initial ground state should have a
non-vanishing overlap with the final
excited state, and the tunneling amplitude is finite\cite{Ruzin}.

The density of states is directly related to the amplitude of the tunneling
process for a semi-infinite
Wigner crystal, after an interstitial is introduced at its edge, see
Eq.~(\ref{rho}). In the semiclassical
approximation this amplitude can be expressed in terms of the action
accumulated along the optimal classical
trajectory,
\begin{equation}
\langle\psi(x,t)\psi^\dagger(x,0)\rangle\sim\exp[-S(t)].
\label{amplitude}
\end{equation}
As we now show,
the slow stage of tunneling contributes
a logarithmically
divergent contribution\cite{Larkin} as $|t|\to\infty$. This divergence
justifies the use of a
(\ref{Lagrangiantheta}) harmonic in deformations, and also enables one to
treat the low-energy
tunneling semiclassically, not only in the limit $g\ll1$, but at {\sl any}
value of $g<1$.

Let us consider the evolution of the right half of the pinned Wigner
crystal, $x>0$. To connect the initial
local deformation $\tilde{\rho} (x,t=0)=2\delta (x)$ with the final relaxed
state
$\tilde{\rho}(x,t=\infty)=0$, the optimal trajectory must run in
imaginary time,
$t=i\tau$. The appropriate solution of the equations of motion gives
\begin{equation}
\tilde{\rho} (x,\tau)=\frac{2}{\pi}\frac{v\tau}{x^2+(v\tau)^2},\quad x>0.
\label{instanton}
\end{equation} This solution is applicable for sufficiently large
$x^2+(v\tau)^2$, so that the harmonic
description, see Eqs.~(\ref{Hamiltonian1}), (\ref{Lagrangiantheta}) is
valid. Note that the dimensionless
displacement at the origin, $\theta (x=+0,t)=-\pi$, remains constant in
time, which allows us to use
consistently the limit of a $\delta$-function in the initial condition.
Eq.~(\ref{instanton}) describes the spread and relaxation of the
deformation created by the tunnelling electron.
The associated potential energy,
$V_{\rm def}(\tau)$, decreases monotonically with time.
Specifically, with
the help of (\ref{instanton}), one has:
\begin{equation}
V_{\rm def}(\tau)=\frac{v}{2\pi g}\int_0^\tau dx (\partial_x\theta)^2=
                     \frac{2 v}{\pi g}\int_0^\tau dx
\frac{(v\tau)^2}{\left[x^2+(v\tau)^2\right]^2},
\label{Vdef}
\end{equation}
which gives $V_{\rm def}(\tau) = 1/(2g\tau)$.
The kinetic energy is equal to the potential energy (virial theorem), and
thus we find
\begin{equation}
S(\tau)=2\int_{\omega_D^{-1}}^{\tau}V_{\rm def}(\tau)=\frac{1}{
g}\ln(\omega_D\tau)
\label{Sdef}
\end{equation}
for the tunneling action.  Here we have used $1/\omega_D$ as the
short-time cut-off for the
slow part of the evolution.
Finally, upon using Eqs.~(\ref{rho})  we can extract
the tunnelling density of states,
$\rho
(E)\propto E^{\alpha -1}$ with $\alpha=1/g$.

\section{Electron transport in a Luttinger liquid with a barrier}
\label{transport}

In the preceding Section we obtained the tunnelling density of states for
adding an electron at
energy $E$ into a  Luttinger liquid.  We considered two cases: Tunnelling
into an infinite
Luttinger liquid and tunnelling into the end of a semi-infinite system.  In
both cases the DOS
vanished as a power law of energy for an electron gas with repulsive
interactions. Here we
consider an infinitely long interacting electron gas with a single defect
or barrier, localized at
the origin. Of interest is electron transport through the barrier. We first
consider two limiting
cases, a very large barrier in Subsection A, and a very small barrier in
Subsection B. In
Subsection C we show how the crossover between these two limits can be
understood by
considering general barrier strengths, but weak electron interactions. A
general picture of the
crossover is discussed in Subsection D. Finally, we consider the special
case of resonant
tunneling in Subsection E.

\subsection{Large Barrier}
\label{strong}

An infinitely high barrier breaks an electron gas into two de-coupled
semi-infinite pieces.  Each piece
can be described by either of the quadratic Lagrangians,
(\ref{Lagrangiantheta}) or
(\ref{Lagrangianphi}). For a very high, but finite barrier, we can consider
electron tunnelling from one
semi-infinite piece into the other as a perturbation.  The appropriate
tunnelling term to add to the
Hamiltonian, is of the form
\begin{equation}
H_{\rm tun} = t_0 [ \psi_1^\dagger (x=0) \psi_2(x=0) + h.c.],
\label{htun}
\end{equation}
where $\psi_1$ ($\psi_2$) is the electron operator in the left (right)
semi-infinite Luttinger liquid.  Here
$t_0$ denotes the (bare) tunnelling amplitude.  Using (\ref{bosons}) these
operators can be readily
expressed in terms of the boson fields, $\theta$ and $\phi$.  However, with
an infinitely high barrier the
displacement field $\theta(x)$ is pinned at the origin, so one can take
$\psi(x=0) = \exp[i\phi(x=0)]$.

The two-terminal conductance through the point contact can now be computed
perturbatively for small
tunneling amplitude $t_0$.  In the presence of a voltage $V$ across the
junction,  the tunneling rate to
leading order can be obtained from Fermi's Golden rule:
\begin{equation}
I = {2\pi e \over \hbar} \sum_n s_n |\langle n| H_{\rm tun}  |0\rangle |^2
           \delta(E_n - E_0 - s_n eV).
\label{goldenrule}
\end{equation}
The sum on $n$ is over many-body states in which an electron has been
transferred across the junction
in the $s_n = \pm 1$ direction.   It is straightforward to re-express this as
\begin{equation}
I = {e t^2_0\over 2\pi\hbar}
\int dE \left[ \rho_1^>(E) \rho_2^<(E-eV) - \rho_1^<(E-eV) \rho_2^>(E) \right]
\end{equation}
where $\rho_a^>$  ($\rho_a^<$)  is the tunneling densities of states for
adding (removing) an electron at
energy $E$.  These are related by $\rho^<(E) = \rho^>(-E)$.

Upon using the expression (\ref{tunndos1}) for the tunnelling DOS into the
end of a semi-infinite
Luttinger liquid, one readily obtains,
\begin{equation}
I \propto t^2_0 |V|^{(2/g)-2} V,\quad G(V)\equiv\frac{dI}{dV}\propto t^2_0
|V|^{(2/g)-2}.
\label{tuniv}
\end{equation}
For repulsive interactions ($g<1)$, the linear conductance is strictly
zero! This is a simple reflection of the
suppressed density of states in a Luttinger liquid. When $g=1$ a linear
$I-V$ curve is predicted,
consistent with expectations for non-interacting electrons which are
partially transmitted through a
barrier. At finite temperatures the density of states is sampled at $E
\approx kT$, and a non-zero (linear)
conductance is expected.  Generalizing Fermi's Golden rule to $T \ne 0$
gives the expected result for the
(linear response) conductance:
\begin{equation}
G(T) \propto t^2_0 T^{(2/g)-2}.
\label{tung}
\end{equation}

It is instructive to re-cast this result in the language of the
renormalization group (RG).  Specifically, the
vanishing conductance for $g<1$ indicates that the tunneling perturbation,
$t_0$, is {\it irrelevant}.  The
RG can be implemented using the Bosonized representation, in which the
tunnelling term takes the form,
\begin{equation}
S_{\rm tun} = t_0\int d\tau e^{i\phi(\tau)} ,
\end{equation}
with $\phi (\tau) = \phi_1(x=0,\tau) - \phi_2(x=0,\tau)$.
This tunnelling term is aded to the quadratic Lagrangians
(\ref{Lagrangianphi}) for the two semi-infinite
Luttinger liquids.   Since the perturbation $t_0$ acts at a single space
point, $x=0$, it is useful to
imagine ``integrating out" the fields $\phi_a(x)$ for $x$ away from the
origin, leaving only the time
dependence, $\phi_a(x=0,\tau)$. The RG then proceed as follows. In the
frequency domain,
$\phi(\omega)$, one integrates over modes in a shell $\omega_c/b < \omega <
\omega_c$, with
$\omega_c$ a high frequency cutoff of order the Fermi energy. This can be
done by splitting the field
into ``slow" and ``fast" modes, below and inside the shell, respectively:
$\phi = \phi_s + \phi_f$.  To
lowest order in $t_0$ one must average over the fast modes:
\begin{eqnarray}
\langle e^{i\phi} \rangle_f =&  e^{i\phi_s} \langle e^{i\phi_f}\rangle_f
\nonumber\\
 =&  b^{-\Delta} e^{i\phi_s}.
\end{eqnarray}
Here $\Delta$ is the scaling dimension of the operator $e^{ i\phi}$.  The
scaling dimension is most easily
deduced  from the two point correlation function,
\begin{equation}
\langle e^{i\phi(\tau)} e^{-i\phi(0)}\rangle \propto |\tau|^{-2\Delta}.
\end{equation}
>From (\ref{amplitude}) and (\ref{Sdef}) one obtains $\Delta = 1/g$. The RG
transformation is completed by rescaling time, $\tau' = \tau/b$, to restore the
cutoff to $\omega_c$. The resulting action is then equivalent to the original
one with $t_0$ replaced by $t'_0 = t_0 b^{1-\Delta}$.  Upon setting $b=e^\ell$,
and denoting the renormalized tunneling amplitude as $t(\ell)$, one thereby
obtains the leading order differential RG flow equation,
\begin{equation}
{dt\over{d\ell}} = (1-\Delta)t ,
\label{RGt}
\end{equation}
with $\Delta =1/g$.  The perturbative results (\ref{tuniv}) and (\ref{tung})
can be obtained by integrating this RG flow equation until the cutoff is of
order $kT$ (or $eV$), giving
$t_{\rm eff} \sim t_0 T^{(1/g)-1}$ and $G \sim t_{\rm eff}^2$.

For a channel of finite length $L$
coupled to Fermi liquid leads, renormalization in (\ref{RGt}) should be
stopped at the level spacing, i.e., at $e^{\ell}\sim k_FL$. The  conductance
is temperature and voltage independent for  $T, eV\ll \hbar v_{F}/L$.

\subsection{Weak Backscattering Limit}
\label{weak}
Having established that the conductance of a repulsively interacting
Luttinger liquid vanishes at
$T=0$ for weak tunneling , we now turn to the opposite limit in which the
barrier is very weak.
In this limit we can treat the barrier as a small perturbation on an ideal
Luttinger liquid.  The
Lagrangian (\ref{Limp}) is a particularly convenient representation, with
the barrier strength $u$
assumed small. As discussed in Section IIB,  the perturbation proportional
to $u$ backscatters
fractionally charged quasiparticles between the de-coupled right and left
moving Luttinger modes.
What effect does this weak backscattering have on transport?

Consider first a simple RG. As descibed in the previous section, to leading
order in $u$ the RG flow
equation is,
\begin{equation}
{du\over{d\ell}} = (1-\Delta)u,
\label{RGr}
\end{equation}
with the scaling dimension $\Delta$ defined via the correlation function,
\begin{equation}
\langle e^{i2\theta(x=0,\tau)} e^{-i2\theta(x=0,0)}\rangle \propto
|\tau|^{-2\Delta}.
\end{equation}
Evaluating this using the quadratic part of the Lagrangian (\ref{Limp})
gives $\Delta = g$.  With
repulsive interactions ($g<1$) the backscattering strength grows at low
energies. Cutting off the flow equations when the temperature $T$ is
comparable
to the cutoff $\omega_c$, gives an effective backscattering stength
diverging at low temperatures as $u_{\rm eff} \sim u T^{g-1}$. Eventually, at
very low temperatures
the backscattering becomes sufficiently strong that treating it as a
perturbation is no longer
valid.  Nevertheless, one expects that upon cooling the backscattering
strength will continue to
grow, until the system scales into the large barrier regime discussed in
Subsection A.  This implies
that at $T=0$ even a very small barrier will be inpenetrable, and
effectively break the Luttinger
liquid into two decoupled pieces.

It is also possible to directly calculate the conductance through a weak
barrier, perturbatively in
$u$. To extract a two-terminal conductance for an infinite wire (ignoring
Fermi liquid
leads), one can reason as follows. With no barrier present, the right and
left moving Luttinger
modes differ in potential by the bias voltage $eV$. This results in a
transport current  $I=g
(e^2/h) V$. Quasiparticle backscattering between the two modes will tend to
reduce this current.
The reduction can be computed perturbatively using Fermi's Golden rule as
in Subsection A, but
with two differences. First, the charge $e$ in (\ref{goldenrule}) must be
replaced by the
quasiparticle charge, $e^* = ge$. Second, the electron tunneling operator
(\ref{htun}) must be
replaced by the quasiparticle tunneling term proportional to $u$.  At zero
temperature
one obtains,
\begin{equation}
I_{\rm back} \propto u^2 |V|^{2g-2} V,
\end{equation}
as could have been anticipated from the RG flow equation. Notice that $g$
has been replaced by
$1/g$ in going from the strong to weak barrier result.  Likewise, at
temperature $T$, the
backscattering contribution to the (linear) conductance is given by
\begin{equation}
G - g {e^2\over h} \propto - u^2 T^{2g - 2}.
\label{Gweak}
\end{equation}

\subsection{The limit $1-g\ll 1$ }
\label{smallglimit}

We have seen in Section \ref{weak} that
backscattering off a barrier is enhanced dramatically
when the electron gas is repulsively interacting,
with $g<1$.
Indeed,
even a weak scatterer is expected to cause
full reflection at zero
excitation energy. In this section, we show how this surprising result
can be understood in terms of
the physical electrons.
Specifically, we show that
the scattering rate for an electron off an impurity is renormalized by the
electron-electron
interaction,
due to the formation of a Friedel oscillation near the barrier.  This
results in singular
reflection amplitude at $q=2k_F$.
By considering the limit of very weak interaction, $1-g\ll1$,
it is possible to treat the
scattering for {\it arbitrary} barrier strength\cite{Yue}.
This allows us to describe the crossover between
the perturbative results for large and small barrier described above.

Consider first a 1d gas of spinless non-interacting electrons
scattering on a potential $u(x)$ localized near the origin. For simplicity,
we assume the
barrier is symmetric, $u(x)=u(-x)$. This allows us to use the same transmission
and reflection
amplitudes, $t_0$ and $r_0$, to describe both sets of asymptotic wave
functions far from the
barrier:
\begin{equation}
 \phi_{k}(x)  = \frac{1}{\sqrt{2\pi}} \begin{cases}
  {                 e^{ikx}+ r_{0} e^{-ikx},      &    $x < 0$,\cr
                   t_{0} e^{ikx},                &   $x > 0$,}
                   \end{cases}
\label{waveleft}
\end{equation}
for the states incoming from the left, and
\begin{equation}
 \phi_{-k} (x) = \frac{1}{\sqrt{2\pi}}\begin{cases}
 {t_{0} e^{-ikx},                &    $x<0$,\cr
e^{-ikx}+ r_{0} e^{ikx},      &    $x > 0$,}
\end{cases}
\label{waveright}
\end{equation}
for the states incoming from the right. The wave vector $k$ is defined to
be positive.

Scattering from the barrier is modified by electron-electron
interactions,
and can be considered perturbatively in the interaction strength.
To lowest-order we
neglect inelastic
processes in which electrons above the Fermi level lose coherence by
exciting electron-hole pairs.
Within this Hartree-Fock approximation, the many-body electron state can be
described
by a Slater
determinant of single-electron wave functions. Each electron is affected
by an extra average
potential produced by other electrons in the Fermi sea. This potential and
the barrier potential,
$u(x)$, act together as an effective barrier for electron scattering.   The
single-electron wave functions can be found as a solution of the
Schr\"{o}dinger equation with this effective barrier.  Then the transmission
coefficient can  be calculated.

The extra potential consists of two parts: the Hartree potential $V_{H}(x)$
determined by the
electron   density  in the system, and a non-local exchange potential
$V_{\rm ex}(x,y)$
which accounts for Fermi statistics of the electrons.
Within this Hartree-Fock approach,
the single-electron wave
functions  can be found by
the Green's function method. The equation for a single-electron Hartree-Fock
state $\psi_{k}$ is:
\begin{equation}
\psi_{k}(x) = \phi_{k}(x) + \int_{y,z} G_{k}(x,y)  K(y,z)  \psi_{k}(z) ,
\label{hf}
\end{equation}
with $K(y,z) = V_{\rm H}(z) \delta(y-z) + V_{\rm ex}(y,z)$.
The Hartree and exchange potentials are defined as:
\begin{equation}
 V_{H} (x) = \int dy V(x-y) n(y),
\label{hartree}
\end{equation}
\begin{equation}
{V_{\rm ex}}(x,y) =
- {V}(x-y) \sum_{|q|<k_{F}} \psi^{*}_{q}(y) \psi_{q}(x),
\label{fock}
\end{equation}
where $V(x-y)$ is the electron-electron interaction potential, and $n(y)
=\sum_{|q|<k_{F}}
|\psi_{q}(y)|^{2}$ is the electron density.

In a first-order Born approximation, $\psi_{k}$ on the right-hand side of
Eq.~(\ref{hf}) and in Eqs.~(\ref{hartree}) and (\ref{fock}) is replaced by the
unperturbed wave function $\phi_k$.  The unperturbed electron density has
the form:
\begin{equation}
n(x)=\begin{cases}{
   n_{0} + \frac1\pi\int_{0}^{k_{F}} dk  {\rm Re}
  \{ r_0 e^ {-2ikx}\}, & $x<0$,\cr
   n_{0} + \frac1\pi\int_{0}^{k_{F}} dk  {\rm Re}
   \{ r_0{*} e^ {-2ikx}\},& $x>0$.
   }\end{cases}
\label{density}
\end{equation}
>From Eq.~(\ref{density}), at large distances $|x|\gg k_F^{-1}$ the
disturbance of density $\delta
n(x)=n(x)-\rho_0$ caused by a symmetric  barrier decays as
\begin{equation}
\delta n(x) \simeq \frac{|r_0|}{2\pi|x|}\sin(2k_F|x| + \arg r_0).
\end{equation}
It follows from Eq.~(\ref{hartree}) that the oscillations of density
(\ref{density}) produce an
oscillating Hartree potential - commonly referred to as a {\em
Friedel oscillation}. In
contrast to the 3d case, where the density oscillation a distance $R$ from
an impurity decays
as
$1/R^{3}$], in 1d  it decays
only as $1/|x|$.
As we shall see, the
$\pi/k_F$ periodicity and slow decay of the Hartree potential,
gives a contribution to $t_0$ and
$r_0$ which is
logarithmically divergent as $k\to k_F$.

To extract the contribution to $t_0$, consider an
incoming wave from the left
with wave vector $k$.
The modified  wave function $\psi_{k}(x)$  must have  the following asymptotics:
\begin{equation}
\psi_{k}(x) \simeq\frac{1}{\sqrt{2\pi}}
 t_{k}e^{ikx},\quad x\to +\infty,
\label{newwave}
\end{equation}
where ${t}_{k}$ is the modified transmission amplitude.
Thus to find the correction to the transmission amplitude, we only need the
asymptotic form of the Green function $G_{k}(x,y)$ as $x\to +\infty$.
Calculating it with free wave functions, we find:
\begin{equation}
G_{k}(x,y) = \frac{1}{i v_{k}} \begin{cases}{
t_{0} e^{ik(x-y)},                 &  $y<0$,\cr
e^{ik(x-y)} + r_{0} e^ {ik(x+y)},  & $y>0$,}
\end{cases}
\label{green}
\end{equation}
where $v_k$ is the velocity of an electron with wave vector $k$.

The transmission amplitude resulting from Eq.~(\ref{hf}) is then,
\begin{equation}
{t}_{k} = t_0 - \gamma t_0|r_0|^2
\ln\left|\frac{1} {( k- k_{F})d}\right|,
\label{dt}
\end{equation}
where $d$ is a characteristic spatial scale of the interaction potential
$V(x)$ (if the range of the
interactions is shorter than the Fermi wave length, then $d$ should be
replaced by $2\pi/k_F$).
Here $\gamma$ is a dimensionless parameter that characterizes the
strength of the
interaction:
\begin{equation}
\gamma = \frac{V(0)-V(2k_F)}{2\pi v_F}   ,
\label{alpha12}
\end{equation}
with $V(q)$ denoting the Fourier transformation of the interaction.

The zero-momentum contribution $V(0)$ originates from the  exchange term,
whereas $V(2k_{F})$ comes from the Hartree term.
Notice that $\gamma$ vanishes for a contact interaction,
for which $V(q)$ is a constant.
Due to Pauli exclusion, such an interaction can play no role.
For a finite range repulsive interaction, $\gamma$ is positive,
and the transmission is suppressed,
in accord with
the results
of Sections~\ref{strong},\ref{weak}. Indeed, we can relate $g$ and $\gamma$
by calculating the
lowest-order interaction correction to the compressibility of a 1d Fermi
gas. For weak interaction
we find
\begin{equation}
1-g\approx \frac{1}{g}-1\approx\gamma .
\label{gapprox}
\end{equation}
Thus we see that Eq.~(\ref{dt}), upon taking the
proper limit ($|t_0|\ll 1$ or $|r_0|\ll 1$), can be viewed as the
solution of the lowest order RG
equation (Eq.~(\ref{RGt}) or Eq.~(\ref{RGr})respectively).

The first-order result (\ref{dt}) for the transmission amplitude consists
of two equal
contributions.  The first one
corresponds to a plane wave
coming from the left and reflected by the barrier with amplitude $r_0$. It
is  then scattered back
to the barrier by the Friedel oscillation on the left-hand side with
amplitude   $-\frac12\gamma r_{0}^*\ln(1/|k-k_{F}|d)$. Finally
the electron penetrates the barrier with amplitude $t_0$.  The
second contribution is the
product of the amplitudes of the following processes: an electron first
penetrates the barrier with
amplitude $t_0$, then it is reflected back to the barrier by the Friedel
oscillation on  the
right-hand side with amplitude $-\frac12\gamma r_{0}^{*}\ln(1/|k-k_{F}|d)$,
and eventually
reflected by the barrier to the right with amplitude $r_0$. The total
first-order contribution to the
transmission amplitude is the sum of these two coherent
processes.

The result (\ref{dt}) has a logarithmic divergence as $k\to
k_{F}$, no matter how small
the coupling constants (\ref{alpha12}). This
indicates the inadequacy of the first-order  calculation at small
$|k-k_{F}|$.
The second order contribution can be extracted by using
the first-order $\psi_{k}(x)$ (\ref{newwave}) as
a new wave function in
the right-hand side of Eq.~(\ref{hf}), and repeating the previous calculation.
The second order contribution to $t_k$ in (\ref{dt}) takes the form,
\newcommand{\kar} {\gamma \ln\left|\frac{1}{(k-k_{F})d}
\right|}
\begin{equation}
t_{k}^{(2)}= -
\frac12 t_{0}|r_0|^2(2|t_0|^2-|r_0|^2) \left[\kar\right]^{2}.
\label{secondorder}
\end {equation}
Here we have only kept the most divergent terms,
which at second order has the form $[\gamma\ln(1/|k-k_{F}|d)]^2$.
At $n$-th order we expect
terms of the form $[\gamma\ln(1/|k-k_{F}|d)]^n$.
Since all these terms are divergent, a straightforward
perturbative approach is clearly inadequate.
Instead, we adopt a renormalization group
approach, as described below.

The Hartree and exchange potentials depend  on the reflection amplitudes,
and they are modified
along with these amplitudes. In a region $(-l,l)$ close to the origin, the
electrons are scattered by
the bare barrier with transmission amplitude $t_0$ and produce an extra
potential that is
proportional to $|r_0|$. Perturbative calculation for the transmission
amplitude is carried out with
the bare  amplitudes. Such a calculation is justified as long as the
correction from the
perturbative calculation is indeed small.  This is true for $l$
not too large, such that
$\alpha\ln(l/d)\ll 1$.  Beyond this distance, the whole region $(-l,l)$
enclosed   should be
considered as an effective barrier to the electrons outside. This effective
barrier is characterized
by the now renormalized amplitudes, denoted $r$ and $t$.  With these new
amplitudes, we can find the
new Hartree and exchange potentials in the outer region. Then the
perturbative calculation can
be carried out to a larger spatial scale.  To ensure that
perturbation theory is valid in
every step, the above renormalization procedure is done repeatedly for
larger and larger scales.

This idea leads to the following formulation.   We start
with a region of length $2l$
centered around the barrier. The scale $l$ is chosen to be
much larger than $d$
but not too large, so that $1\ll\ln(l/d)\ll \gamma^{-1}$. The modified
transmission amplitude due
to the electron-electron interaction in the region $(-l,l)$ can then be
found by perturbation theory,
\begin{equation}
t_1=t_0 -\gamma t_{0}(1-|t_0|^2)\ell,
\end{equation}
with $\ell=\ln l/d\gg 1$.

We then go to a larger scale, taking the region $(-l,l)$ as a composite
scatterer. Using  the
renormalized transmission amplitude  and correspondingly  renormalizing the
additional Hartree
and exchange potentials,   we repeat the calculation for this
next scale $l\exp(\ell)$.
Then to the next larger scale, which is $l\exp(2\ell)$, and so on.
In general, the iterative
renormalization of the transmission amplitude after $n$ steps of scaling to
larger distances can be
found from
\begin{equation}
t_{n+1}=t_{n} -\gamma t_{n}(1-|t_{n}|^2)\ell.
\label{realrg}
\end{equation}
This  iteration procedure should be stopped at a length scale $1/|k-k_{F}|$,
beyond which the
scattered electron loses phase coherence with the Friedel oscillation, and
the transmission
amplitude is not renormalized any further.

In the continuous limit, Eq.~(\ref{realrg}) becomes
\begin{equation}
\frac{dt}{d\ell}=-\gamma t(1-|t|^2),
\label{realRG}
\end{equation}
where $\ell$ is the logarithm of the length scale. Integrating
equation (\ref{realRG})
from $\ell=0$ to $\ell=\ln (1/|k-k_{F}|)$ and using the
boundary condition
$t|_{\ell =0}=t_0$, we find
a renormalized transmission  amplitude
\begin{equation}
t_k = \frac{t_{0}|(k-k_{F})d|^{\gamma}}{\sqrt{
|r_0|^2 +|t_0|^2|(k-k_{F})d|^{2\gamma}}}.
\label{realrgresult}
\end{equation}
The transmission coefficient ${\cal T}=|t|^2$ is then
\begin{equation}
{{\cal T}}(\epsilon) = \frac{{{\cal T}}_0 |E/D_0|^{2\gamma}}
                    {{\cal R}_0+{{\cal T}}_0|E/D_0|^{2\gamma}},
\label{trenormed}
\end{equation}
where ${\cal T}_{0}= 1-{\cal R}_{0}=|t_0|^2$ is the bare transmission
coefficient, and $D_0=v_F/d$.
The expansion of Eq.~(\ref{realrgresult}) up to the second order in
$\gamma$ coincides with
(\ref{secondorder}).

The renormalized transmission coefficient  (\ref{trenormed}) allows us
to find the temperature dependence of the linear conductance of a 1d
spinless interacting
electron system with a single barrier. At high temperatures $T  > D_0$ the
conductance is given by
the Landauer formula for an ideal Fermi  gas, $G_0=(e^2/2\pi\hbar){\cal
T}_0$. At lower
temperatures the  transmission coefficient is renormalized.  Because of the
smearing of
the Fermi surface, $E$ in Eq.~(\ref{trenormed}) should be replaced by $T$,
and the following
temperature dependence of the linear conductance is  found:
\begin{equation}
 G({T}) = \frac{e^2}{2\pi\hbar}
        \frac{{{\cal T}}_0 (T/D_0)^{2\gamma}}
                    {{\cal R}_0+{{\cal T}}_0(T/D_0)^{2\gamma}}.
\label{conductance}
\end{equation}
Formula (\ref{conductance}) describes explicitly the crossover between the
limits of weak
reflection and weak tunneling considered above in the framework of
Luttinger liquid theory. The
proper expansion of Eq.~(\ref{conductance}) and the use of
Eq.~(\ref{gapprox}) yields the
limiting results, Eqs.~(\ref{tung}) and (\ref{Gweak}). The crossover
temperature depends on the
strength of the barrier, $T^*=D_0({{\cal R}}_0/{{\cal T}}_0)^{1/2\gamma}$.
The differential conductance $G(V)$ at a high voltage $eV > T$ may be
obtained by substitution ${T}\to eV$.

\subsection{Crossover between the weak and strong backscattering}
\label{crossover}

The preceding results can now be pieced together to form a global picture
of the behavior of a scattering defect in a Luttinger liquid.
The perturbative results in IV.A and IV.B describe the
stability of two renormalization group fixed points.
For repulsive interactions ($g<1$), the ``perfectly insulating"
fixed point, with zero electron tunneling $t_0=0$,
is stable, whereas the ``perfectly conducting"
fixed point,
with zero backscattering $u=0$, is unstable.
(For {\it attractive} interactions, $g>1$, the stability
conditions are reversed.)
Provided these are the only two fixed points,
it follows that the RG flows
out of the conducting fixed point eventually make their way
to the insulating fixed point.
In the limit
of very weak repulsive interactions, $1-g<<1$,
we showed this crossover explicitly in Section IV.C,
but it is true more generally.
This is a very striking conclusion,
since it implies that a Luttinger liquid
with arbitrarily weak scatterer, with amplitude $u$,
will cause the conductance to vanish
completely
at zero temperature.  Of course, for $u$ very small, very low
temperatures would be necessary
to see this.  In this scenario, the conductance as a function
of temperature will behave
as follows.  At high temperatures, the system
does not have ``time" to flow out of the perturbative regime, so
the conductance is given by $G \approx (ge^2/h) - u^2 T^{2g - 2}$.
As the temperature is lowered below a scale
$T^* \propto u^{1/(1-g)}$, perturbation theory breaks down.
Eventually, the system crosses over into a low temperature
regime in which the conductance vanishes as $T^{(2/g)-2}$.

The validity of this scenario rests on the assumption that
no other fixed points intervene.
This assumption has been verified both by
quantum Monte Carlo simulations\cite{Moon}, and more recently by
exact non-perturbative methods based on
the thermodynamic Bethe ansatz\cite{Fendley}.

\subsection{Resonant tunnelling}
\label{resonant}

We now briefly consider the phenomena of resonant tunnelling
through a double barrier in a 1d Luttinger liquid.
Our reasons are two fold: (i) Experiments on quantum Hall edge
states, discussed in the next Section, have measured resonances which can
be compared with Luttinger liquid theory, (ii) The nature of the
crossover between the weak and strong backscattering limit, determines
the lineshape and temperature dependence of resonance tunnelling
peaks in a Luttinger liquid.

As a point of reference, we first review resonant tunneling theory
for a 1d non-interacting electron gas.
Consider then 1d electrons incident on a double barrier structure,
with a (quasi-) localized state between the barriers.
As the chemical potential $\mu$ of the incident
electron sweeps through the
energy of the localized state, $\epsilon_0$, the conductance
will exhibit a peak described by,
\begin{equation}
G = {e^2\over h} \int d\epsilon f'(\epsilon-\mu)
{\Gamma_L \Gamma_R \over {(\epsilon-\epsilon_0)^2 + \Gamma^2}} .
\label{Breit}
\end{equation}
Here $\Gamma_L$ and $\Gamma_R$ are tunneling rates from the resonant (localized)
state to the left and right leads and
$\Gamma = (\Gamma_L+\Gamma_R)/2$.  The Fermi function
is denoted $f(\epsilon)$.  At high temperatures, $T>\Gamma$, the resonance
has an amplitude $\Gamma/T$ and a width $T$.
At low temperatures, the lineshape is Lorentzian, with a temperature
independent width $\Gamma$.  Moreover, when the left and right barriers are
identical, the on-resonance transmission at zero temperature is perfect,
$G=e^2/h$.

How is this modified when the electron gas is an interacting
Luttinger liquid?   Since arbitrarily weak
backscattering causes the zero temperature
conductance to vanish, one might expect that
resonances are simply not present at $T=0$.
As we now show, this is not the case.  Rather, perfect
resonances are possible, but in striking contrast to (\ref{Breit}) for non-
interacting electrons, they become infinitely sharp in the zero
temperature limit.

To see this, it is convenient to consider the limit
of a very weak double-barrier structure.
For non-interacting elecrons, resonant tunneling is not normally studied for
weak backscattering, since in this limit the
transmission is large even off resonance, which tends to
obscure the resonance.
However, in a Luttinger liquid, the off-resonance conductance vanishes
at zero temperature, leaving an unobscured
resonance peak, as we now argue.

Consider then scattering of a pure Luttinger
liquid from a small barrier, denoted $u(x)$, which is non-zero only for
$x$ near zero.
For resonant tunnelling, the potential should
be taken to have a double-barrier structure.
This potential couples to the electron density,
via an additional term in the Hamiltonian,
\begin{equation}
 {\cal H}_{\rm imp} =
\int dx u(x) \psi^\dagger(x)\psi(x).
\end{equation}
This can be re-expressed in terms of the boson fields
by inserting the expression (2.6) for the electron operator.
We assume that $\theta(x)$ is slowly varying on the scale
of the potential, and so replace it by $\theta(x=0)$.
The integration over $x$ can then be performed to give,
\begin{equation}
{\cal H}_{imp} = {1 \over 2} \sum_{n=-\infty}^\infty u_n
e^{i2n \theta(x=0)} ,
\label{Himp1}
\end{equation}
where the coefficients, $u_n = u_{-n}^*={\hat u}(2nk_F)$, are proportional
to the Fourier transform of $u(x)$ at momenta given
by $n$ times $2k_F$. For a symmetric barrier, $u(x) = u(-x)$ the
coefficients $u_n$ are real.

In Equation (\ref{Himp}) we retained only the first term in this sum. The
higher order terms correspond to processes where n-electrons are
simultaneously backscattered, each by a momenta $2k_F$, from one Fermi point
to the other. Alternatively, (\ref{Himp1}) can be viewed as an effective
potential $u_{\rm eff}(\theta (x=0) )$, which is invariant under the
transformation $\theta\rightarrow\theta+\pi$.
Since $\theta/\pi$ is the number of particles to the left of
$x=0$, $u_{\rm eff}$ may be regarded as a weak pinning potential
in the Wigner crystal picture.

As we shall now see, for weak backscattering, the single
electron process, ie. the $2k_F$ backscattering term $u_1$,
is the most important. This follows readily from a perturbative RG calculation,
as in Section~\ref{weak}.  Specifically, the scaling dimension, $\Delta_n$,
of the
perturbation $u_n$, defined via the correlation function,
\begin{equation}
\langle e^{i2n\theta(x=0,\tau)} e^{-i2n\theta(x=0,0)}\rangle \propto
|\tau|^{-2\Delta_n},
\end{equation}
with imaginary time $\tau$, can be readily evaluated
using the quadratic part of the Lagrangian (\ref{Limp}),
to give $\Delta_n = n^2 g$.
Thus the leading order RG flow equations for the renormalized coupling
$u_n(\ell)$ are simply,
\begin{equation}
{du_n\over{d\ell}} = (1-n^2g)u_n .
\end{equation}

With increasing $n$, the scaling dimension increases,
and the perturbation, $u_n$, becomes less relevant (or more
irrelevant), suggesting we need only focus on $u_1$.
However, imagine fine-tuning $u_1$ to zero.  As we shall see,
this corresponds to tuning to a resonance.
The next most important backscattering process is
$u_2$.  But notice, that for $g>1/4$ this backscattering
amplitude scales to zero, as do {\it all} the higher order processes,
$u_n$ with $n>2$.  Thus, {\it if} $u_1$ is tuned to zero,
provided $g>1/4$ one expects {\it perfect} transmission
at $T=0$.  On the other hand, as we have shown earlier
when $u_1$ is nonzero,
it grows under RG (for $g<1$), and the conductance {\it vanishes}
at $T=0$.  Thus at zero temperature, there will
be an {\it infinitely sharp resonance peak}
as $u_1$ is varied through zero!

How easy is it to achieve such a resonance?
For $g>1/4$, the criterion is that the renormalized value of $u_1$ vanishes. In
general, $u_1$  is complex, so that the resonance condition requires the
simultaneous tuning of two parameters.  However, for symmetric barriers,
$u_1$ is real and only a single parameter need be tuned.

A resonance with no width is in striking constrast
to the conventional result for non-interacting electrons.
In that case, the width is set by the tunnelling rate $\Gamma$
from the localized state between the barriers, into
the leads.  The higher the barrier, the smaller
the decay rate, and the narrower the resonance.
An electron in a localized state between two barriers
in a Luttinger liquid will be unable to decay (at $T=0$), since the tunnelling
density of states into the ``leads" vanishes.
The electron remains localized forever, with an infinite ``lifetime" -
which gives a simple explanation of the infinitely sharp resonance.
At a finite temperature, the electron will be able to decay,
since the tunnelling DOS into a Luttinger liquid
is non-zero away from the Fermi energy.  One thus expects that
the resonance will be thermally broadened, as we now confirm.

Consider tuning through such a perfect resonance
by varying a parameter.
It is convenient to denote by $\delta$ the ``distance"
from the peak position in the control parameter.
Close enough to the resonance one has $u_1 \propto \delta$.
For very small $\delta$ the RG flows will thus pass
very near to the
perfectly conducting fixed point, since all of the other
irrelevant operators will scale to zero before
$u_1$ has time to grow large.
Eventually, $u_1$ does grow large and the flows
crossover to the insulating fixed point.
Temperature serves as a cutoff to the RG flows, as usual.
This reasoning reveals that for both $\delta$ and temperature
small, the conductance will depend
only on the universal crossover trajectory which
joins the two fixed points.
The uniqueness of the RG trajectory implies that
the conductance will be described by a universal
crossover scaling function.
Moreover, since $u_1$ grows with exponent $1-g$,
the conductance, which is generally a function of both
$\delta$ and $T$, will only depend on these parameters
in the combination, $\delta/T^{1-g}$.  Thus, for
small $T$ and $\delta$, the resonance
lineshape is given by a universal scaling function,
\begin{equation}
G(T,\delta,g) = {ge^2\over h} \tilde G_g(c_g \delta/T^{1-g}),
\label{universal}
\end{equation}
where $c_g$ is a (non-universal) constant.

The scaling function $G_g(X)$ depends {\it only} on
the interaction parameter $g$, but is otherwise {\it universal},
independent of all details.  The limiting behavior of
$\tilde G_g(X)$, may be deduced from
the perturbative limits.  For small argument $X$, the
perturbation theory result (\ref{Gweak}) implies
\begin{equation}
\tilde G_g(X) = 1 - X^2.
\end{equation}
For large argument, corresponding to the limit $T\rightarrow 0$,
the scaling function must match on to the low temperature regime (\ref{tung}),
which gives a $T^{(2/g)-2}$ dependence.
This implies that for $X\rightarrow\infty$,
\begin{equation}
\tilde G_g(X) \propto X^{-2/g}.
\end{equation}

The scaling form for the conductance near resonance,
reveals that the width of the resonance scales to zero
with a power of temperature, $T^{1-g}$.  Notice that in
the non-interacting limit, $g \rightarrow 1^-$, this reduces to the
expected temperature independent linewidth.  Moreover,
the finite temperature resonance lineshape is
non-Lorentzian, with tails decaying more rapidly, as $X^{-2/g}$.
Again, for $g \rightarrow 1$ this reduces to the expected
Lorentizian form for the non-interacting electron gas.

The exact scaling function which interpolates between the
small and large $X$ limits, can be computed
for $g=1/2$ by re-Fermionizing the Luttinger liquid.
More recently,
Fendley et. al.\cite{Fendley} have computed
$G_g(X)$ for {\em arbitrary} $g$,
using the thermodynamic Bethe Ansatz.

\section{Applications}
\label{applications}

Here we demonstrate two applications of the theory reviewed in the previous
sections. The first one relates the general theory to the experimentally
observable  transport phenomena in a mesoscopic quantum Hall effect. The
second one uses the mathematical tools described above in the theory of
Coulomb blockade.

\subsection{Tunneling between Quantum Hall Edge States}
\label{edgestates}

\subsubsection{Edge states in the regime of quantum Hall effect}

Despite the firm theoretical basis upon which the Luttinger
liquid theory rests, there has been precious little
compelling experimental evidence that real
one-dimensional electron gases are anything but Fermi liquids.
This, despite the fact that in recent years it has become possible
to fabricate single channel quantum wires.
Several recent experiments\cite{Tarucha,Yacoby} have reported interesting
transport data on such quantum wires, and offered
possible interpretations in terms of Luttinger liquid theory.
But the analysis is complicated by several factors.
Firstly, it is difficult to eliminate unwanted impurity
scattering along the wire.  When this is strong,
it causes backscattering and localization, destroying the Luttinger liquid
phase.  But even in a very clean wire, the Luttinger liquid parameter
$g$, which determines the power law of tunnelling density of states,
is unknown, depending on the details of the Coulomb interaction
strength, which complicates the interpretation.
Fortunately, there is another experimental system which is expected
to exhibit Luttinger liquid behavior, and does not suffer from the above
difficulties - namely edge states in the quantum Hall effect.

The quantum Hall effect
occurs at low temperatures in two-dimensional electron gases with low
carrier density, when
placed in a strong perpundicular magnetic field.  The key experimental
signature consists of quantized plateaus in the Hall conductance
as the magnetic field is varied.  In the plateaus, the Hall
conductance, $G_H$, is ``locked" to values which are simple rational numbers
of the quantum conductance,
$G_H = \nu e^2/h$, with rational $\nu$.
In the {\it integer} quantum Hall effect, $\nu$ is an integer,
whereas in the {\it fractional} quantum Hall effect,
$\nu=p/q$ with integer $p$ and odd integer $q$.
In the plateaus the longitudinal conductivity, $\sigma_{xx}$,
vanishes rapidly as $T \rightarrow 0$.

The integer quantum Hall effect (IQHE) can be understood
in terms of 2d non-interacting electrons
in a magnetic field.  When the Fermi energy lies
between Landau levels, there is an energy gap of order
the cyclotron energy, $\omega_c =eB/m$, which accounts
naturally for the vanishing $\sigma_{xx}$.  But at the edges of the sample,
there are gapless current carrying states.  In a semi-classical
picure, these can be thought of as orbits which ``skip" along the edge
due to the magnetic field and edge confining potential.  The
direction of the skipping, clockwise or counter-clockwise, is determined
by the sign of the magnetic field.  In a full quantum treatment,
these edge orbits become quantum modes.  For $n$ full Landau levels
in the bulk, there are $n$ edge modes, one for each Landau level.

The IQHE edge modes are equivalent to the right moving sector
of a 1d non-interacting electron gas.  In a Hall bar geometry,
the modes on the top and bottom edges move in opposite directions.
For $\nu=1$ there is a single right moving
mode on the top edge, and a left mover on the bottom.
These two modes can be described by the Hamiltonian (\ref{lrhamiltonian})
with $g=\nu=1$.  That is, they are mathematically equivalent
to a 1d non-interacting electron gas.  However, since the two modes
are spatially separated - on opposite sides of the Hall bar - impurities
cannot cause backscattering and localization.

Quantization of the Hall conductance can be understood very simply
in terms of the edge states.  As discussed in Section II.A,
raising the chemical potential of the right moving mode with respect
to the leftmover by an amount $\mu$, leads to a transport current,
$I=G\mu$ with $G=ge^2/h$ from (\ref{Landauer}).  In the Hall bar geometry,
$\mu$ corresponds to the transverse (Hall) voltage drop, so $G$
is a Hall conductance - appropriately quantized for $g=1$.

The fractional quantum Hall effect occurs when a single Landau level
is partially filled, with a fractional filling $\nu$.
For non-interacting electron there would be an enormous ground
state degeneracy in this case.  At rational filling fractions,
$\nu=p/q$, the electron Coulomb repulsion lifts this
degeneracy, leading to a unique ground state with an energy
gap to excited states.  The resulting FQHE state is predicted
to have very interesting properties, such as quasiparticle excitations
with fractional charge and {\it statistics}.

In pioneering work, Wen\cite{Wen} argued that FQHE states  should also have gapless
current carrying edge modes. But in contrast to the IQHE, these edge modes
were argued to be Luttinger liquids. Specifically, for the Laughlin sequence of
fractions, at filling $\nu=1/q$ with odd integer $q$, a single chiral Luttinger
liquid mode was predicted. For a Hall bar geometry, as in the IQHE with
$\nu=1$, there would be one right moving mode on the top edge, and one
leftmover on the bottom. Again, these two modes can be described by the
Hamiltonian (\ref{lrhamiltonian}), but now with {\it fractional} Luttinger
liquid parameter, $g = \nu =1/q$.   Remarkably, the Luttinger liquid parameter
for FQHE edge states is {\it universal}, determined completely by the bulk FQHE
state, and independent of all details.

\vspace{-0.5cm}

\begin{figure}
\centerline{
\psfig{file=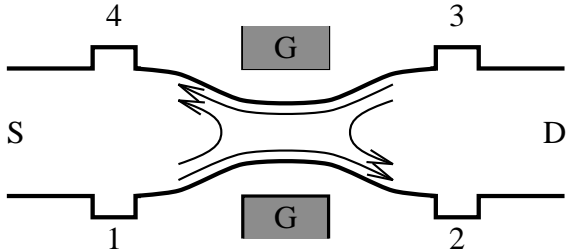,width=3in
}
}
\vspace{0.1in}
\setlength{\columnwidth}{3.2in}
\nopagebreak
\centerline{
\caption{Schematic portrait of a point contact, in which the top and bottom
edges of a Hall fluid are brought together by an electrostatically controlled
gate (G), allowing for the tunneling of charge between the two edges.  Here S
and D denote source and drain, respectively. 
\label{portrait}}}
\end{figure}

For hierarchical FQHE states, at fillings $\nu \ne 1/q$,
multiple edge modes are predicted\cite{Wen}, in some cases moving in both
directions along a given edge.  In the following we will
focus for simplicity on the Laughlin sequence of states, and in
particular on the most robust state at $\nu=1/3$.

Although FQHE edge modes at filling $\nu=1/3$, and a Luttinger liquid
with $g=\nu$ are essentially equivalent, there are some subtle differences
which must be kept in mind.  Specifically, in the Luttinger liquid
the fundamental (right moving) charge $e$ Fermion operator (the electron)
is given from (\ref{bosons}) by $\psi_R \sim  e^{i(\theta + \phi)}$.  This
can be
re-written in terms of the right and left moving Boson modes, $\phi_{R/L} =
g\phi \pm \theta$, as $\psi_R \sim e^{i(2\phi_R + \phi_L)}$.
This operator thus adds charge $2/3$ into the right moving Boson mode,
and $1/3$ into the left.  In the FQHE, this corresponds
to adding $2$ fractionally charged quasiparticles (charge $e/3$)
to the top edge of the Hall bar, and $1$ to the bottom edge.
But this is a {\it non-local} operator, and hence unphysical
in the FQHE, whereas it is nice local operator in a quantum wire.

\subsubsection{Inter-edge tunneling at a Point Contact}

As discussed in Sections~\ref{tunneling} and \ref{transport}, the most
remarkable property of a Luttinger liquid is the vanishing density of states
for tunnelling electrons.  To allow for
intermode edge tunnelling in the FQHE, it is necessary
to bring together the opposite edges of the Hall bar.
This can be achieved by gating the electron gas, as depicted
schematically in the Figure~\ref{portrait}.  

\begin{figure}
\centerline{
\psfig{file=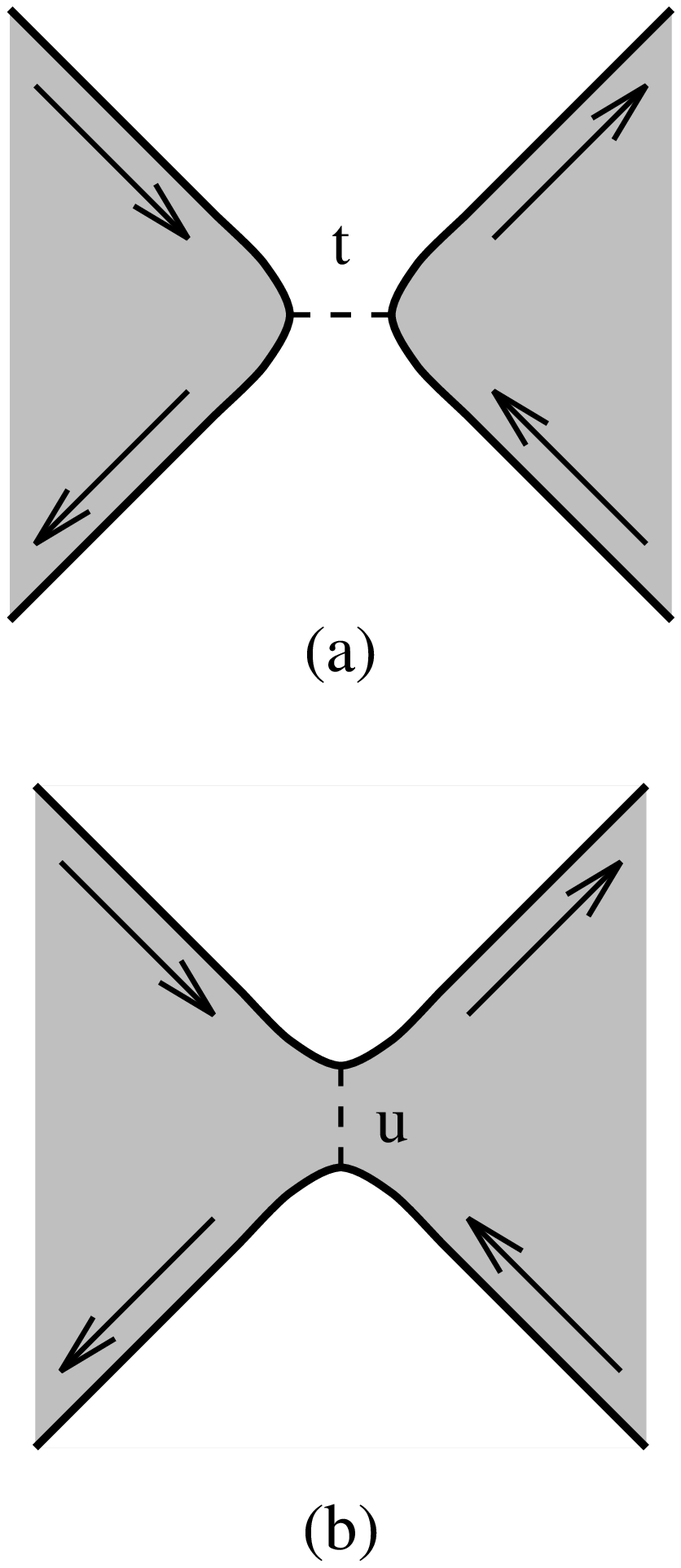,width=3in
}
}
\vspace{0.1in}
\setlength{\columnwidth}{3.2in}
\nopagebreak
\centerline{
\caption{A quantum Hall point contact in the ({\sl a}) weak tunneling
limit and ({\sl b}) the weak backscattering limit.  The shaded regions represent
the quantum Hall fluid with edge states depicted as lines with arrows.  The
dashed line represents a weak tunneling matrix element connecting the two edges.
\label{limits}}}
\end{figure}
Consider specifically, a gate which brings the opposite edge modes together at
one point.  This is equivalent to putting a single defect in an otherwise clean
Luttinger liquid. In this geometry, charge can tunnel between the top and bottom
edge modes, through the narrow strip of FQHE fluid (see Fig.~\ref{portrait}).
The appropriate term to add to the Hamiltonian is given in (\ref{Himpurity}),
and corresponds to the transfer of a fractionally charged Laughlin quasiparticle
(charge $e^* = \nu e = e/3$) from top to bottom, with amplitude $u$. In a
Luttinger liquid, this same perturbation corresponds to a weak $2k_F$ electron
backscattering, with ``backflow".

Since $g=\nu <1$ this weak backscattering amplitude, $u$,
grows at low temperatures,
and the system crosses over into a large barrier regime,
as discussed in Section~\ref{transport}.  The large barrier limit
corresponds to Figure~\ref{limits}{\sl a}, in which the incident
top edge mode is almost completely reflected.
Weak tunnelling from left to right can be treated
perturbatively, in the amplitude, $t_0$.  In this case,
it is an electron which is tunnelling, with charge $e$.
As shown in Section~\ref{strong}, (Eqn. (\ref{tung})), this leads
to a conductance which vanishes as a power of temperature,
\begin{equation}
G \propto t_0^2 T^{(2/\nu)-2} \sim T^4 ,
\label{fqhe}
\end{equation}
for $\nu=1/3$.

\begin{figure}
\centerline{
\psfig{file=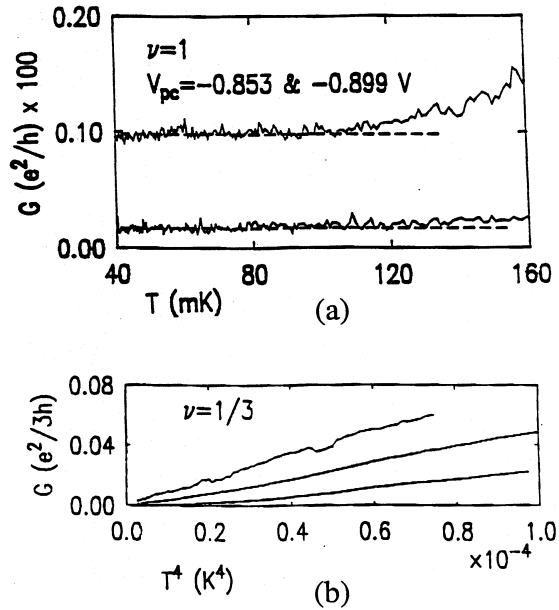,width=3in
}
}
\vspace{0.1in}
\setlength{\columnwidth}{3.2in}
\nopagebreak
\centerline{
\caption{Conductance of a quantum point contact as a function of
temperature for ({\sl a}) $\nu=1$, and ({\sl b}) $\nu = 1/3$. Taken from
Ref.~\protect \onlinecite{Webb}.
\label{gtemp}}}
\end{figure}

Fig.~\ref{gtemp} shows data\cite{Webb} for the conductance as a function of
temperature through a point contact in an IQHE fluid at $\nu=1$ and a  FQHE fluid
at $\nu=1/3$. 
The difference in behavior is remarkable. For $\nu =1$ the
conductance approaches a constant at low temperatures, whereas for $\nu=1/3$ the
conductance continues to decrease upon cooling. Moreover, the low temperature
behavior for $\nu=1/3$ is consistent with the $T^4$ dependence predicted in
(\ref{fqhe}).  This data provides  experimental evidence for the Luttinger
liquid, a phase discussed theoretically over 30 years earlier.

\begin{figure}
\centerline{
\psfig{file=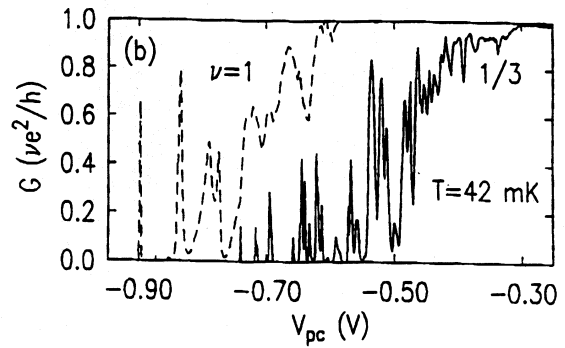,width=3in
}
}
\vspace{0.1in}
\setlength{\columnwidth}{3.2in}
\nopagebreak
\centerline{
\caption{Two-terminal conductance as a function of gate voltage of a GaAs
quantum Hall point contact taken at 42 mK. The two curves are taken at magnetic
fields which correspond to $\nu=1$ and $\nu = 1/3$ plateaus. Taken from
Ref.~\protect \onlinecite{Webb}.
\label{peaks}}}
\end{figure}

In this same experiment, the conductance through the point contact
was measured as a function of a gate voltage, controlling
the pinch-off.  The data in Fig.~\ref{peaks}, shows a sequence
of reproducible conductance peaks. A natural interpretation is
that these are resonant tunnelling peaks, through
a state localized in the vicinity of the
point contact.  As discussed in Section~\ref{resonant},
resonant tunnelling peaks are expected in
a Luttinger liquid whenever the amplitude for the
dominant backscattering process ($u_1$) is tuned to zero.
In the FQHE, this process corresponds to tunnelling a single
$e/3$ quasiparticle from the top to
bottom edge.  Since $\nu>1/4$, the higher order processes
which involve multiple quasiparticle tunnelling are irrelevant.
Thus, for a symmetric scattering potential, finding
a resonance requires tuning only a single parameter, such as the
gate potential.

\vspace{-0.5cm}

\begin{figure}
\centerline{
\psfig{file=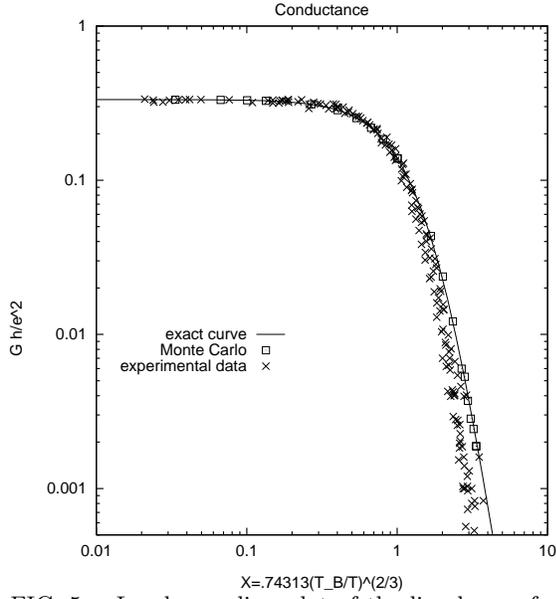,width=3in
}
}
\vspace{0.1in}
\setlength{\columnwidth}{3.2in}
\nopagebreak
\centerline{
\caption{
Log-log scaling plot of the lineshape of resonances at different
temperatures.  The x axis is rescaled by $T^{2/3}$.
The crosses represent experimental data of Ref.~\protect \onlinecite{Webb}
at temperatures between $40mK$ and $140mK$.  The squares are the results
of the Monte Carlo simulation, and the solid line is
the exact solution from Ref. \protect \onlinecite{Fendley}. 
\label{curve}}}
\end{figure}

Fig.~\ref{curve} shows a scaling plot of one of the resonances for
$\nu=1/3$ in Fig.~\ref{peaks} from the data of Webb {\sl et al}. 
The widths of the resonances at several
different temperatures
have been rescaled by $T^{2/3}$, as suggested by (\ref{universal}). Since
the peak heights were
also weakly temperature dependent (and roughly one third of the quantized value
$(1/3)e^2/h$) the amplitudes have also been normalized to have unit height
at the peak. The temperature scaling of the peak widths is indeed very well fit
by $T^{2/3}$.  Also shown in Fig.~\ref{curve} are quantum Monte Carlo data and an
exact computation from Bethe Ansatz for the universal scaling function in
(4.38).  The agreement is striking. Although the experimental lineshape does
drop somewhat faster in the tails, the shape is distinctly non Lorentzian with a
tail decaying with a power close to that predicted by theory.  It should be
emphasized that the experimental data does not represent a ``perfect resonance",
since the peak amplitude is dropping (slowly) upon cooling, rather than
approaching the quantized value, $(1/3)e^2/h$. By varying an additional
parameter besides the gate voltage (such as  the magnetic field) though, it
should be possible to find a perfect resonance for $\nu=1/3$.

\begin{figure}
\centerline{
\psfig{file=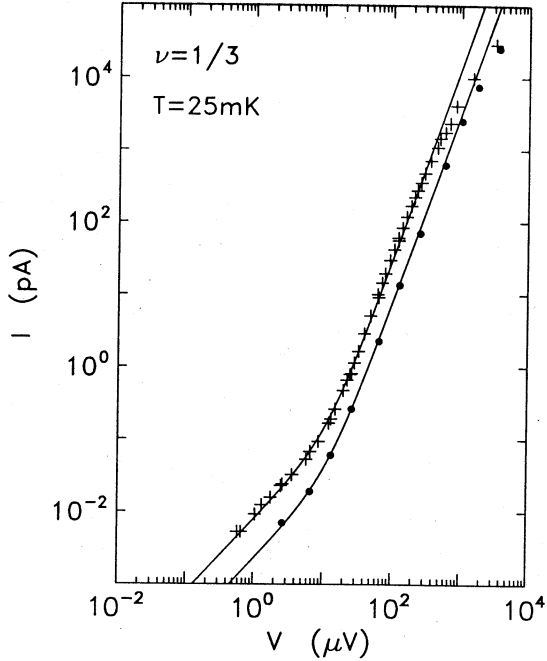,width=3in
}
}
\vspace{0.1in}
\setlength{\columnwidth}{3.2in}
\nopagebreak
\centerline{
\caption{Current-voltage characteristics for tunneling from a bulk-doped
$n+$ GaAs into the edge of a $\nu=1/3$ fractional quantum Hall fluid for two
samples: Sample 1 at magnetic field $B=13.4$T (crosses), and sample 2 at
$B=10.8$T (solid circles). The solid curves represent fits to the theory; the
extracted values of the exponents for the two curves are $\alpha=2.7$ and
$\alpha=2.6$ respectively. (Taken from Ref.~\protect\onlinecite{Chang}.)
\label{changiv}}}
\end{figure}

In a very recent experiment, Chang et. al.\cite{Chang} have succeeded in
tunnelling into a FQHE edge from a bulk metallic system. The 2DEG was exposed,
in situ, by cleaving a bulk GaAs sample.  A thin insulating barrier was grown
onto the cleaved face. Heavily doped material was deposited on top of the
insulating layer.  The experiment measured the tunnelling conductance from the
heavily doped ``metal" through the insulating barrier into the FQHE edge. The
tunnelling conductance from a Fermi liquid into a $g=\nu$ chiral Luttinger
liquid, is predicted to vanish as, $G \sim T^{(1/\nu) -1}$.  Since the DOS in a
Fermi liquid is constant, this power is $1/2$ as large as for tunnelling between
two $g=\nu$ chiral Luttinger liquids. Similarly, at zero temperature, the I-V
curve is predicted to vary as $I \sim V^{1/\nu}$. Figure~\ref{changiv} shows
data for the $I$-$V$ curve which is consistent with this power law.

\vspace{-0.1cm}

\subsection{ Coulomb blockade of strong tunneling}
\label{coulomb}

\vspace{-0.1cm}

\begin{figure}
\centerline{
\psfig{file=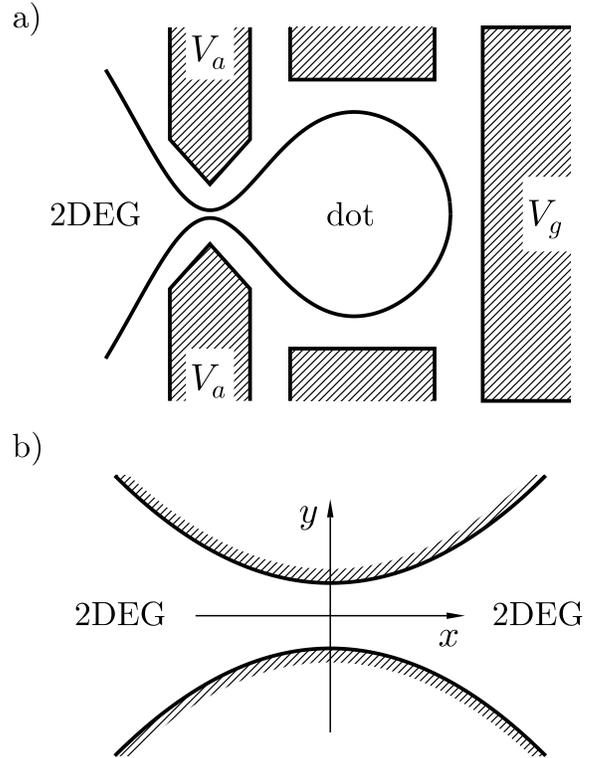,width=3in
}
}
\vspace{0.1in}
\setlength{\columnwidth}{3.2in}
\nopagebreak
\centerline{
\caption{({\sl a}): Metallic grain connected to a lead by a channel (taken from
Ref. \protect \onlinecite{strong}). The charge on the grain is controlled by the
gate voltage, $V_g$, applied to an electrode which is coupled capacitively to the
grain. ({\sl b}): One-dimensional channel connecting the grain to the lead.
\label{dot}}}
\end{figure}

In this section we apply the mathematical tools developed for a 1d Luttinger
liquid, to an apparently unrelated physical problem - namely the Coulomb
blockade. Consider a conducting metallic grain (or ``dot") connected to an
infinite lead by a junction, see Fig.~(\ref{dot}{\sl a}). 

The grain is also coupled capacitively to another gate electrode. The
electrostatic energy of the grain is
\begin{equation}
H_Q=\frac{(Q-eN)^2}{2C_0},
\label{charging}
\end{equation}
where $Q$ is the charge of the grain and $C_0$ is its capacitance.
The parameter $N$ is proportional to the gate voltage $V_g$. Variation of the
gate voltage $V_g$ changes the equilibrium charge of the grain. If the
channel to the lead is almost depleted, and its conductance $G\to 0$, then
the charge $Q$ can only take on discrete values $Q={{\cal N}}e$. The equilibrium
charge $\langle Q\rangle$ should minimize the energy (\ref{charging}) within
the set of integer ${\cal N}$, and therefore changes
in a stepwise fashion with the
gate voltage, $\langle Q\rangle = {\rm Int} Ne$. On the other hand, if $G$ is
large, one expects the grain charge to vary linearly with the gate
voltage, $\langle Q\rangle =Ne$.
Of interest is the behavior in the crossover region
between these two limiting cases, as the conductance to the
lead is varied.

This problem is non-trivial, because the
electron-electron interaction term (\ref{charging})
must be treated non-perturbatively.
As we shall see, if the junction is modelled
as a single-mode channel with tunable reflection
amplitude $r$,
it is possible\cite{Flensberg,strong} to apply the methods described in
Section~\ref{bosonization}.
Indeed, using bosonization, Matveev\cite{strong} has
shown that $\langle Q\rangle =Ne$ at $r=0$, and found analytically the
function $q(N)\equiv\langle Q\rangle - Ne$ for weak reflection. (Here
and after, $\langle\dots\rangle$ denotes a ground-state
average.) The average charge is directly related to the ground state energy,
$q(N)=-(C_0/e)\partial E/\partial N$.  Below we
follow
closely the line of arguments presented in Ref.~\onlinecite{strong}.

To find the ground state energy $E$, we need to supplement the
Hamiltonian (\ref{charging}) by a part describing non-interacting
electrons. For the case of a grain with size $L$ large compared to
the Bohr radius, $a_B=\hbar^2/me^2$, this can be simplified. If
the motion within the grain is chaotic, then the dwelling time for an electron
entering the grain through a single-mode channel is $\tau_d\sim
\delta^{-1}$, where $\delta$ is the electron level spacing in the grain. On the
other hand, the quantum charge fluctuations which destroy the
discreteness of $\langle Q\rangle$ occur on the time scale
$\tau_q\sim 1/E_C$, where $E_C\sim e^2/C_0$ is the single-electron charging
energy.  The latter time scale is relatively short, $\tau_q/\tau_d\sim
a_B/L\ll 1$ (the estimate is performed for a two-dimensional grain). Over
this time, electrons participating in the charge fluctuations ({\sl i.e.}, those
that pass through the channel) do not  get back into the channel after
``exploring'' the grain. It means that we need to add to Eq.~(\ref{charging}) a
Hamiltonian describing only the electrons moving through a
one-dimensional channel. In the absense of a barrier within the channel,
this Hamiltonian is
\begin{equation}
H_0=iv_F\int dx
              \left[\psi_R^{\dag} (x)\partial_x\psi_R (x) -
                      \psi_L^{\dag} (x) \partial_x\psi_L (x)\right],
\label{channel1}
\end{equation}
where $\psi_R^{\dag} (x)$ and $\psi_L^{\dag} (x)$ are the creation operators for
right- and left-movers, respectively. We have used here a linearized electron
spectrum, since the energy
scale $E_C$ is much smaller than the Fermi energy,
$E_C/E_F\sim
(a_B/L)(k_Fa_B)^{-2}\ll 1$.

The definition of the charge operator $Q$ in Eq.~(\ref{charging}) depends on
where we ``draw" the boundary between the channel and the grain.
This arbitrariness in convention only affects the phase of the
oscillations of $q(N)$.
But this phase
does not carry any physical meaning,
provided the gate voltage is small
enough not to deplete the grain entirely. Therefore, $Q$ can be
viewed as the charge passed {\sl e.g.} through the middle ($x=0$) of the
channel, Fig.~(\ref{dot}b) into
the dot,
\begin{equation}
 Q=e\int_0^{\infty} dx
\left[\psi_R^{\dag} (x)\psi_R (x) +
                      \psi_L^{\dag} (x) \psi_L (x)\right].
\label{charge1}
\end{equation}

Now, with the help of  Eq.~(\ref{smoothpart}), we can bosonize the
Hamiltonian (\ref{channel1}), which gives
Eq.~(\ref{Hamiltonian1}) with $g=1$,
\begin{equation}
H_0 = {v_F \over {2\pi}} [(\partial_x \phi)^2 + (\partial_x\theta)^2] .
\label{channel2}
\end{equation}
Bosonization of the interaction part (\ref{charging}) is also straightforward,
after we put
$\psi_R^{\dag} (x)\psi_R (x) +\psi_L^{\dag} (x) \psi_L (x)
\to -\partial_x\theta (x)/\pi$ into (\ref{charge1}),
\begin{equation}
H_C=E_C\left[\frac{1}{\pi}\theta (0) - N\right]^2, \quad E_C=\frac{e^2}{2C}.
\label{chargingb}
\end{equation}

In the absence of a barrier in the channel, it is easy to show that
$q(N)=0$.
Indeed,
after the transformation
\begin{equation}
\theta (x)\to\theta (x) +\pi N,
\label{shift}
\end{equation}
the Hamiltonian $H_0+H_C$ does not depend on $N$; there is no Coulomb
blockade at $r=0$.  The ground state energy starts to depend on $N$, if a
barrier causes {\em backscattering} in the channel. The corresponding
Hamiltonian (cf Eq.~(\ref{Himp})) in the boson variables is
\begin{equation}
H_{\rm imp}=-\frac{u}{\pi v_F}D\cos[2\theta (0)],
\label{backscattering}
\end{equation}
where $u$ is the $2k_F$--component of the scattering potential, and $D$ is
a high energy cut-off. The transformation (\ref{shift}), makes evident the
periodic dependence of the full Hamiltonian, $H_0+H_{\rm imp}+H_C$, on
$N$.

To find the correction to
the ground state energy to first order
in $|r|$, denoted $E_1$y,
the perturbation (\ref{backscattering}) must be averaged
in the ground state of
the Hamiltonian $H_0+H_C$, Eqs.~(\ref{channel2}), (\ref{chargingb}).
It then follows directly from Eqs.~(\ref{channel2}), (\ref{chargingb}) and
(\ref{shift}), that $E_1\propto \cos (2\pi N)$. The proportionality coefficient
can be estimated with the help of the RG equation (\ref{RGr}). In the absence of
the local interaction (\ref{chargingb}), the renormalization (\ref{RGr}) is
valid
for all energy scales and can be taken to the limit $\ell\to\infty$, which
results in a vanishing effective scattering potential, $E_1=0$. With a finite
$E_C$,
the renormalization should be stopped at $\exp(-\ell)\sim E_C/D$, which at
$g=1$ leads to $E_1=-{\rm const}\cdot(u/v_F)E_C\cos 2\pi N$.
To leading order in the backscattering,
$|r|=u/v_F$, so that
\begin{equation}
E_1=-{\rm const}\cdot |r|E_C\cos 2\pi N.
\label{almost}
\end{equation}
An exact result for $E_1$ obtained\cite{strong} in the framework of the
Debye-Waller theory for the quadratic Hamiltonian  $H_0+H_C$, yields ${\rm
const}=e^{\bf C}/\pi^2$, with ${\bf C}\approx 0.5772$ being Euler's
constant. Finally,
\begin{equation}
\frac{q(N)}{e}=-\frac{e^{\bf C}}{\pi}|r|\sin 2\pi N.
\end{equation}

\section{Conclusion}
In this paper we have reviewed recent theoretical results for
transport in a one-dimensional (1d) Luttinger liquid.
For simplicity, we have ignored electron spin, and
focussed exclusively on the case of a single-mode.
Moreover, we have considered only the effects of a single
(or perhaps several) spatially localized impurities.
Even with these restrictions, the predicted transport
behavior is qualitatively different than for a non-interacting
electron gas.  Specifically,
for repulsive interactions, even a weak impurity potential
causes complete backscattering, and the conductance
{\it vanishes} completely at zero temperature.
This can be understood in terms of the
vanishing density of states, to tunnel an electron throught the barrier
from one semi-infinite Luttinger liquid to another.
At finite temperature the tunnelling electron
has finite energy, and the conductance is non-zero,
varying as a power of temperature.
The precise power law depends on the dimensionless
parameter (conductance) $g$ which characterizes the Luttinger liquid.

For a very weak barrier, at elevated temperatures,
the backscattering is weak, but grows rapidly with cooling.
This growth can be understood physically in terms of the fact that
the discrete backscattered charge in this limit,
is {\it less} than the electron charge $e$,
but rather given by $ge$ with $g<1$ for a repulsively interacting electron gas
(see Eqn. (2.22)).
For inter-edge tunnelling in the fractional quantum Hall effect,
this process corresponds to
the backscattering of a Laughlin quasiparticle
(with $g=\nu = 1/3$, say), but such fractionally
charged excitations are in fact a generic property of the
1d Luttinger liquid, and should be present,
for example, in narrow quantum wires.
The fractional charge might be directly measureable
via shot-noise experiments through a single impurity.

The transport behavior in a 1d Luttinger liquids is of course
much richer once one relaxes the restrictions of
a single impurity and a single channel.
The case of a single channel with many impurities has been considered
by a number of authors\cite{Giamarchi}.  In the absence of electron
interactions,
all states are localized in 1d, and the system is an insulator.
With repulsive interactions, localization effects are enhanced,
and the system is insulating, even though the notion of
single particle localized eigenstates is no longer operative.
However, for spinless electrons with sufficiently
strong {\it attractive} interactions, the system is predicted
to undergo an insulator-to-metal transition.
In the metallic phase, the {\it conductivity} is predicted to
diverge as a power law of temperature (with a variable power
greater than one), in contrast to a normal 3d metal,
in which there is a finite residual resistivity at $T=0$.
While the electron interaction
in a quantum wire is clearly repulsive,
a long skinny gate across a fractional quantum Hall
fluid at, {\sl e.g.}, filling $\nu=1/3$, creates a system which
is isomorphic to a (spinless) 1d electron gas with strong {\it attractive}
interactions\cite{Renn}.
By varying the gate potential along such a line junction,
it should be possible to tune through this 1d localization transition.

For a real one-channel quantum wire, it is of course necessary
to take into account the spin of the electron\cite{Yue,long}.
The method of bosonization can readily accomodate the spin degree
of freedom.  The spinful interacting electron gas
has two modes, one which carries the charge and the other the
spin\cite{Solyom,Mahan}. These two modes will generally propagate at {\it
different} velocities, a phenomena known as charge/spin separation.  The effects
of  impurity scattering on the spinful electron gas are qualitatively
similar to that without spin.  Specifically,
with repulsive interactions, a single impurity will typically
be sufficient to
completely backscatter both the charge and spin modes.
The conductance will be driven to zero
as a power law of temperature, with possible
logarithmic corrections.  Resonant tunnelling
is also possible for a spinful electron gas, but can occur in several
different guises, depending on the charge and spin state
of the localized state.

In a wider quantum wire, several transverse modes will co-exist at the Fermi
energy. This can also be treated with bosonization. Generally, associated with
each channel is one gapless charge mode and one gapless spin mode. The
tunnelling density of states to add an electron, will generally still be
singular, as for a single channel, but the associated exponents become smaller as
the number of channels increase\cite{Matveev,multichannel}. Multiple channels are
also present at the edges of hierarchical fractional quantum hall
states\cite{hierarchy}, such as at filling $\nu=2/3$ . Multi-component models
are also necessary to treat  charge fluctuations under the Coulomb blockade
conditions (Section~\ref{coulomb}) in the realistic case of electrons with a
spin degree of freedom. Indeed, a recent theory of Coulomb blockade effects for
coupled quantum dots\cite{harold}, has yielded predictions which
are in quantitative accord with experiment\cite{Waugh}.

\section{Acknowledgements}

We have both benefitted from extensive interactions
with numerous friends and collaborators
over the past years.  We are particurly indebted
to Igor Aleiner, Harold Baranger,
Albert Chang, Steve Girvin,
Bertrand Halperin,
Charlie Kane, Anatoly Larkin, Patrick Lee, Hsiuhau Lin, Andreas Ludwig,
Allan MacDonald,
Konstantin Matveev, Kyungson Moon,  Joe Polchinski, Igor Ruzin,
Boris Shklovskii, M. Stone, Richard Webb and Hangmo Yi.
We gratefully acknowledge the National Science
Foundation for support under grants DMR-9400142, PHY89-04035
and DMR-9528578 (University of
California at Santa Barbara), and No. DMR-9423244 (University of Minnesota).

\end{multicols}

\vfill\eject
\narrowtext

\end{document}